\documentclass[aps,prb,manuscript]{revtex4}
\usepackage{amsmath,amssymb,graphicx,color}

\begin{document}
\draft
\title{Tunneling of Hybridized Pairs of Electrons through a One-Dimensional Channel}

\author{Godfrey Gumbs$^1$, Danhong Huang$^{2}$,  Julie Hon$^1$, M. Pepper$^{3,4}$, and Sanjeev Kumar$^{3,4}$      }
\affiliation{$^{1}$Department of Physics and Astronomy, Hunter College of the
City University of New York, 695 Park Avenue, New York, New York 10065, USA\\
$^2$Air Force Research Laboratory, Space Vehicles Directorate, Kirtland Air Force Base, New Mexico 87117, USA\\
$^3$London Centre for Nanotechnology, 17-19 Gordon Street, London, WC1H 0AH, United Kingdom\\
$^4$Department of Electronic and Electrical Engineering, University College London, Torrington Place, London, WC1E 7JE, United Kingdom}

\date{\today}

\begin{abstract}
Recently, the electron transport through a quasi-one dimensional (quasi-1D) electron gas was
investigated  experimentally as a function of the confining potential. We present a physical model for quantum
ballistic transport of electrons through a short conduction channel,
and investigate the role played by the Coulomb interaction in modifying
the energy levels of two-electron states at low temperatures as the width of the channel
is increased. In this regime, the effect of the Coulomb interaction
on the two-electron states has been shown to lead to four split energy
levels, including two anti-crossings and  two crossing-level states.
Due to the interplay between the anti-crossing and crossing of the energy levels,
the ground state for the two-electron model switches from one anti-crossing
state for strong confinement to a crossing state for intermediate confinement
as the channel width is first increased, and then returned to its original
anti-crossing state. This switching behavior is related to the triplet spin degeneracy as well as the Coulomb repulsion and reflected in the ballistic
conductance. Here, many-body effects can still affect
electron occupations in the calculation of quantum ballistic conductance
although it cannot vary the center-of-mass velocity.
\end{abstract}

\maketitle

\section{Introduction}
\label{sec:1}

In this paper, we review the importance of many-body effects on the ballistic electron
transport in a quasi-one-dimensional (1D) electron gas  with varied confinement potential.
This area of research has been receiving a considerable amount of attention
in recent times ever since it was discovered that for a range of  electron
distribution and potential strength, the ground state of a 1D
quantum wire splits into two rows with a Wigner lattice beginning to form.
It was also demonstrated  that when a perpendicular magnetic field is applied,
a double-row electron formation  may change completely  into a single row due
to an enhanced  confinement potential. Furthermore, it has been  verified experimentally
that weak confinement, in competition with the electron-electron interaction, causes the electron
level occupation to reorder so that the  ground state, conforming to the standard
or common type, passes through the  excited levels. The
data in Ref.\,[\onlinecite{pepper}] show that the energy levels may be controlled by exploiting
their separate geometric dependence on  confinement and electron density. This means
that in simulating the electron transport data, many-body effects must be considered.
\medskip
\par

It has been  well known that electrostatic potential confinement of a two-dimensional
electron gas (2DEG) used to create a quasi-1D wire\,\cite{add1} gives rise to quantization
of the conductance\,\cite{add2,add3} in integer multiples of $2e^2/h$ which is not affected
by a weak electron-electron interaction\,\cite{add4}. The long-range Coulomb interaction
between electrons becomes relatively important at low electron densities resulting in the
formation of a 1D Wigner crystal\,\cite{add1,add5,add6}. But, the role played by the
Coulomb repulsion between electrons is also made greater until it overcomes the
confinement potential, as the density is  increased, at which point the ground
state and one of the excited states are interchanged\,\cite{add7} which may
result in hybridization and anti-crossing. In Refs.\,[\onlinecite{pepper,paper1}],
conductance measurements were reported for weakly confined quantum wires
in a 2DEG and determined by the boundaries of top split-gates.
\medskip
\par

Experiment has shown that  making the confinement potential less effective
results in the appearance of two rows, accompanied by a sudden change in
conductance $G$ from zero to $4e^2/h$. This behavior may be attributed to
the possibility that there was no coupling between these two rows so that
each row contributes independently and additively. Another way to account for
this is to say that their energy eigenstates become hybridized and the resulting state
causes a breakdown of the single-particle picture when the Coulomb interaction
becomes important. Recent investigations have confirmed that there exists
a Coulomb interaction  between the rows  resulting in this anomalous jump in the
conductance\,\cite{add10}.
\medskip
\par

The devices used in Refs.\,[\onlinecite{pepper}] and [\onlinecite{paper1}] were fabricated using electron
beam lithography on $300\,$nm deep GaAs/AlGaAs heterostructures.
Typically, the sample consisted of split gates, $\sim 0.4\,\mu$m long and $0.7\sim 1.0\,\mu$m
wide, and a top gate of width $\sim 1.0\,\mu$m defined above the split
gates, separated by a $200\,$nm layer of cross-linked polymethyl methacrylate.
After partial illumination, the carrier density
and mobility were estimated to be $\sim 1.5\times 10^{11}\,$cm$^{-2}$ and
$\sim 1.3\sim 3\times 10^6\,$cm$^2/$Vs, respectively. The two-terminal conductance,
$G=dI/dV$, was measured at $70\,$mK, using a $77\,$Hz voltage of $10\,\mu$V.
Previously, the conductance through two laterally aligned but uncoupled
parallel wires formed by surface gates have been shown to be the sum of the conductances
of each individual wire, resulting in plateaus at multiples of $4e^2/h$,\,\cite{add11,add12}
indicating that hybridization of states within a wire is a many-body effect but not
a single-particle one. Two side-by-side wires with very small inter-wire separation have
lent support for the theory, i.e., there exists a coupling between the parallel wires\,\cite{add13,add14}.
When this coupling between wires becomes strong, the electron wave functions hybridize, forming
bonding and antibonding states, which manifest as anticrossings in the 1D energy subbands.
Our model calculations\,\cite{DAN} further confirm that the minimum energy gap between the
states occurs at the point of anticrossing but is not given by the energy difference between
the symmetric and antisymmetric states.
\medskip
\par

\begin{figure}[t]
\centering
\includegraphics[width=0.65\textwidth]{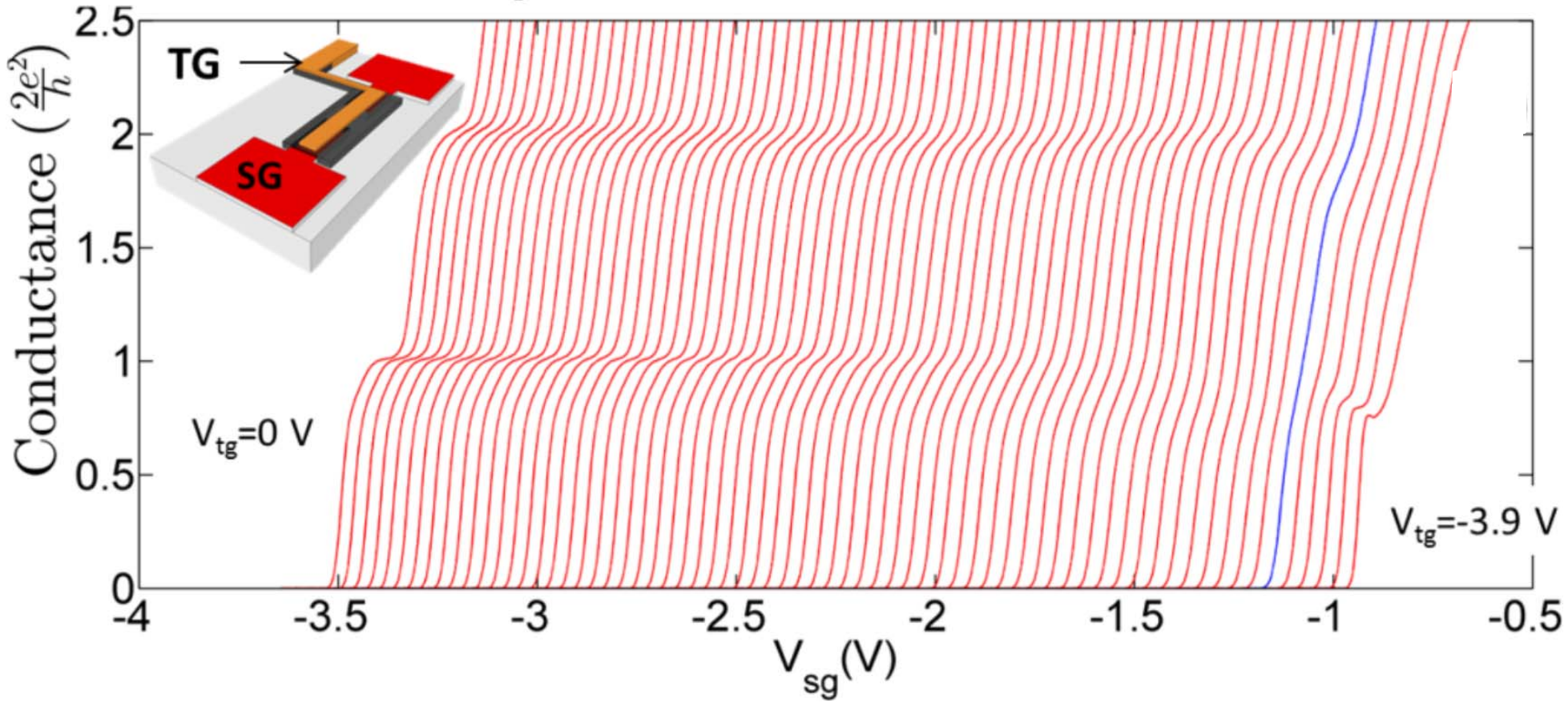}
\caption{(Color online) Schematic illustration of a device used in our experiments (left-upper corner), including a pair of split gates and a top gate. Additionally, typical measured conductance features
of the device are also shown as a function of split-gate voltage, $V_{\rm sg}$ for various fixed top-gate voltages, $V_{\rm tg}$.}
\label{FIG:1}
\end{figure}

In Fig.\, \ref{FIG:1}, we present a schematic illustration of a device used in the experiments
carried out in Refs.\,[\onlinecite{pepper,paper1}], showing a pair of split gates and a
top gate which adjust the confinement potential and carrier density
by choosing their voltages suitably. Figure\ \ref{FIG:1} also shows typical conductance features
obtained with the device used in Ref.\,[\onlinecite{pepper}] as a function of the split-gate voltage,
$V_{\rm sg}$, for fixed top-gate voltage, $V_{\rm tg}$. The traces for strong
confinement are on the left-hand-side,  whereas those for weak confinement are on the right.
When the confinement is weak, the $2e^2/h$ step may be lost and the $4e^2/h$ appears
as the lowest plateau\,\cite{pepper}. But, as the confinement is reduced further, the
$2e^2/h$ plateau is found to be restored.
\medskip
\par

With regard to the interpretation that the carriers  separate  into two rows, the  observed
emergence of the crossing or anticrossing of energy levels needs explanation, preferably
with the use of a quantum-mechanical theory. This may be verified by calculations
of the kinetic, direct Coulomb and exchange energies of electrons in wires with intermediate widths
as well as for two extreme limits of very narrow and wide wires. In fact, we have recently
demonstrated that these cases may be tracked down to the physical mechanism responsible for
switching of the ground state as the wire width is varied from one value to another\,\cite{DAN}.
In the presence of the Coulomb interaction, two-electron states may be employed as a basis set for
constructing the anticrossing-level states\,\cite{pepper} when two electrons travel ballistically
along a quasi-1D channel. The corresponding calculations have shown that the significance of the Coulomb
induced level anticrossing within a quantum wire may be adjusted by varying the
confinement potential with a top gate voltage\,\cite{DAN}.
\medskip
\par

There has  been related work on conductance measurements  of a quasi-1D wire
having a quantum dot within the channel due to the presence of an
impurity, as well as imperfections in the device geometry.\,\cite{paper5}
These undesired features may  result in  results  differing from integer multiples
of $2e^2/h$ for the conductance steps\,\cite{half-step}or oscillations superimposed on the conductance
trace\,\cite{paper4}. Electron tunneling through the quantum dot in the channel
as well as interference effects due to electron back-scattering from an impurity potential
are believed to be responsible for these deviations in the values of the conductance
plateaus of narrow quantum wires.\,\cite{paper4}
\medskip
\par

In the next section, we present a theoretical approach
for calculating the conductance for  a quasi-1D quantum wire
at a low density of electrons.  For this,
we calculate the lowest energy eigenstates for a pair of interacting
electrons  within a  confinement region. We explicitly determine  the
ground state of a dilute electron liquid and consequently the lowest
conductance plateau. The complex two-electron tunneling\,\cite{pair,pair1}
is not  included in this review since it does not contribute to the formation
of conductance-plateaus.  Furthermore, we highlight below
that there is a range of wire widths for which  two-electron transport
is mediated by anticrossing level states based on the Coulomb interaction, and,
therefore, it is not possible to describe the conductance by using a single-particle formalism.

\section{Theoretical Formulation of the Problem}
\label{sec:2}

We will exploit the results for the eigenstates of a pair of interacting
electrons within a harmonic confining potential\,\cite{Wagner,Bryant}.
In Ref.\,[\onlinecite{Wagner}], a symmetric harmonic potential  was introduced.
According to Kohn's theorem\,\cite{Bryant}, for this potential the Coulomb
interaction should only affect the relative motion of electrons but not that for the center-of-mass.
It has been pointed out that as a perpendicular magnetic field is increased,
the ground state will oscillate between a spin singlet and a spin triplet.
Bryant\,\cite{Bryant} showed these electron correlation effects depend on the area of
containment. By solving the Schr\"odinger equation exactly
for two interacting electrons, it becomes clear how correlations
may select the ground state and give rise to quasiparticles which
participate in the transport.
\medskip
\par

Coherent wavefunctions of two interacting electrons may be maintained during
their transport along the channel if scattering by randomly distributed impurities
and defects (negligible lattice scattering at low temperatures) is very small
for high-mobility short channel samples. Also, if the transmission coefficient
for two injected electrons is almost perfect and the inelastic scattering between
different two-electron states is nearly vanishing, we are able to use a
quantum ballistic transport model for two interacting electrons then the Coulomb interaction
between electrons in the channel can be fully taken into account. Ballistic transport of
two-electron clusters is assumed along the channel ($y$) direction. However, the finite width of
a conduction channel in the transverse $x$ direction gives rise to quantization of the split cluster
energy levels. Each level is  assigned to have a free-electron-like kinetic energy for a ballistically moving
noninteracting two-electron cluster. This leads to a hierarchy of subbands with quadratic
wave-vector dependence. The cluster energy levels are given by
$E^{(p)}_{j,k_y}=E^{(p)}_{j}+\hbar^2k_y^2/m^\ast$ with label $p$ and
$j=1,\,2,\,3,\,4$, where $k_y$ is the wave vector of electrons along the channel,
$m^\ast$ is the electron effective mass and the ``subband edges'' $E^{(p)}_{j}$
are presented after Eq.\,(\ref{add16}) in terms of the dimensionless Coulomb integrals.
For a fixed linear electron density $n_{\rm 1D}$, the two-electron chemical potential
$\mu_{\rm p}(T,n_{\rm 1D})$ within the channel may be calculated using

\begin{equation}
n_{\rm 1D}=\frac{2}{\pi}\,\sum_{j=1}^{4}\,\int\limits_{0}^\infty
dk_y\left[\exp\left(\frac{E^{(p)}_{j,k_y}-\mu_{\rm
p}}{k_{\rm B}T}\right)+1\right]^{-1}\ ,
\label{e2}
\end{equation}
which is expressed in terms of the temperature $T$ of the system. Also, the chemical
potentials for the left ($L$) and right ($R$) electrodes (with areal electron density
$n_{\rm 2D}$) are $\mu_{\rm L}^{(p)}(V_b,\,T,n_{\rm 2D})=\mu_p(T,n_{\rm 1D})+eV_b$ and
$\mu_{\rm R}^{(p)}(V_b,\,T,n_{\rm 2D})=\mu_p(T,n_{\rm 1D})-eV_b$, respectively,
in the presence of a low biased voltage $V_b$.
\medskip
\par

For quantum ballistic charge/heat transport of two interacting electrons in the channel,
the charge $(\alpha =0)$ and  heat $(\alpha =1)$ current densities are calculated
from\,\cite{lyo5}

\begin{equation}
J^{(\alpha)}(V_b,\,T,\,n_{1D})=\frac{(-2e)^{1-\alpha}}{\pi}\,\sum_{j=1}^{4}\,\int\limits_{0}^\infty dk_y\,
(E_{j,k_y}^{(p)}-\mu_{\rm p})^\alpha
\left|v_{j,k_y}\right|
\left[f_{\rm L}(E^{(p)}_{j,k_y})-f_{\rm R}(E^{(p)}_{j,k_y})\right]\ ,
\label{e1}
\end{equation}
where $v_{j,k_y}=\hbar k_y /m^\ast$ is the half of the
total group velocity of the two-electron state, $f_{\rm L}(E^{(p)}_{j,k_y})$ and
$f_{\rm R}(E^{(p)}_{j,k_y})$ are the Fermi functions for noninteracting
two-electron states in the left ($L$) and right ($R$) electrodes with chemical potentials
$\mu_{\rm L}^{(p)}$	and $\mu_{\rm R}^{(p)}$, respectively, for noninteracting two-electron states.
\medskip
\par

In formulating our theory,  we start by considering two interacting electrons in an elongated
quantum dot within a quantum wire. For this quantum dot, an anisotropic confinement is assumed with a shorter confining
length $\xi _x$ across the channel ($x$ direction) with a longer confining length $\xi_y$ along the channel
($y$ direction). Here, the Hamiltonian for two interacting electrons can be written as

\begin{eqnarray}
\hat{\cal H}({\bf r}_1,\,{\bf r}_2)&=&
\hat{\cal H}_0({\bf r}_1)+\hat{\cal H}_0({\bf r}_2)+{\cal U}_{\rm C}(|{\bf r}_1-{\bf r}_2|)
\nonumber\\
\hat{\cal H}_0({\bf r}_i)&=&\frac{\hat{\bf p}_i^2}{2m^\ast}+{\cal V}_{\rm c}({\bf r}_i)\ ,\ \ \ \
{\cal U}_{\rm C}(|{\bf r}_1-{\bf r}_2|)=\frac{e^2}{4\pi\epsilon_0\epsilon_{\rm r}|{\bf r}_1-{\bf r}_2|}\ ,
\label{add1}
\end{eqnarray}
where $i=1,\,2$ labels each  electron, $\hat{\bf p}_i=
-i\hbar\mbox{\boldmath$\nabla$}_{{\bf r}_i}$, ${\cal V}_{\rm c}({\bf r}_i)$ is the confining
potential for the conduction channel,
$\hat{\cal H}_0({\bf r}_i)$ is the single-electron Hamiltonian and
${\cal U}_{\rm C}(|{\bf r}_1-{\bf r}_2|)$
represents the electron-electron interaction in a medium with background
dielectric constant $\epsilon_{\rm r}$.
\medskip
\par

The single-particle eigenstates $\phi_{\alpha}({\bf r})\equiv\langle{\bf r}|\phi_\alpha\rangle$
can be determined from the Schr\"odinger equation
$\hat{\cal H}_0({\bf r})\,\phi_{\alpha}({\bf r})=\varepsilon_{\alpha}\,\phi_{\alpha}({\bf r})$,
where the eigenfunctions $\{\phi_{\alpha}({\bf r})\}$ constitute a complete
orthonormal set $|\phi_\alpha\rangle$ in the single particle Hilbert space.
After properly anti-symmetrizing the two particle basis, including both the orbital and spin parts, we obtain

\begin{eqnarray}
&&\Psi_{\alpha_m,\beta_n}({\bf r}_1,s_1;{\bf r}_2,s_2)
\nonumber\\
&=& \frac{1}{\sqrt{2}}\left[\phi_{\alpha_m}({\bf r}_1)\chi_m(s_1)\,\phi_{\beta_n}({\bf r}_2)\chi_n(s_2)-
\phi_{\beta_n}({\bf r}_1)\chi_n(s_1)\,\phi_{\alpha_m}({\bf r}_2)\chi_m(s_2)\right]\ ,
\label{add2}
\end{eqnarray}
where $\chi_m(s_i)$ is the spinor for the spin state of an electron. Here, the basis states in Eq.\,(\ref{add2})
are degenerate eigenstates of the noninteracting Hamiltonian
$\hat{\cal H}_0({\bf r}_1,\,{\bf r}_2)\equiv
\hat{\cal H}_0({\bf r}_1)+\hat{\cal H}_0({\bf r}_2)$ with
$\hat{\cal H}_0({\bf r}_1,\,{\bf r}_2)\,\Psi_{\alpha_m,\beta_n}({\bf r}_1,s_1;{\bf r}_2,s_2)=
\big(\varepsilon_{\alpha_m}+\varepsilon_{\beta_n}\big)\,\Psi_{\alpha_m,\beta_n}({\bf r}_1,s_1;{\bf r}_2,s_2)$.
We restrict ourselves in the following to the case when  only the lowest two orbitals $\alpha_m=\alpha$
and $\beta_n=\beta$ are populated.
\medskip
\par

If we assume that two electrons stay in the same spin state with
$\chi_1=\chi_2=|\uparrow\rangle$, the spinor part can be factored out, giving rise to

\begin{equation}
\Psi({\bf r}_1,s_1;{\bf r}_2,s_2)=\frac{1}{\sqrt{2}}
\left[\phi_{\alpha}({\bf r}_1)\,\phi_{\beta}({\bf r}_2)-
\phi_{\beta}({\bf r}_1)\,\phi_{\alpha}({\bf r}_2)\right]\chi_1(s_1)\chi_1(s_2)\ .
\label{add3}
\end{equation}
However, if we assume that two electrons remain in opposite spin state with
$\chi_1=|\uparrow\rangle$ and $\chi_2=|\downarrow\rangle$, then the result becomes

\begin{equation}
\Psi({\bf r}_1,s_1;{\bf r}_2,s_2)=\frac{1}{\sqrt{2}}\left[
\phi_{\alpha}({\bf r}_1)\chi_1(s_1)\,\phi_{\beta}({\bf r}_2)
\chi_2(s_2)-\phi_{\beta}({\bf r}_1)\chi_2(s_1)\,
\phi_{\alpha}({\bf r}_2)\chi_1(s_2)\right]\ .
\label{add4}
\end{equation}
Consequently, the subspace of the lowest states for two independent electrons can be
spanned by the following basis\,\cite{GG}

\begin{eqnarray}
|\Psi_{1}({\bf r}_1,s_1;{\bf r}_2,s_2)\rangle&=&\frac{1}{\sqrt{2}}\,
\phi_{\alpha}({\bf r}_1)\phi_{\alpha}({\bf r}_2)
\left(|\uparrow\rangle_1|\downarrow\rangle_2-
|\downarrow\rangle_1|\uparrow\rangle_2\right)\ ,
\nonumber\\
|\Psi_{2}({\bf r}_1,s_1;{\bf r}_2,s_2)\rangle&=&\frac{1}{\sqrt{2}}\,
\phi_{\beta}({\bf r}_1)\phi_{\beta}({\bf r}_2)
\left(|\uparrow\rangle_1|\downarrow\rangle_2-
|\downarrow\rangle_1|\uparrow\rangle_2\right)\ ,
\nonumber\\
|\Psi_{3}({\bf r}_1,s_1;{\bf r}_2,s_2)\rangle&=&\frac{1}{2}\,
\left[\phi_{\alpha}({\bf r}_1)\phi_{\beta}({\bf r}_2)+
\phi_{\beta}({\bf r}_1)\phi_{\alpha}({\bf r}_2)\right]
\left(|\uparrow\rangle_1|\downarrow\rangle_2-
|\downarrow\rangle_1|\uparrow\rangle_2\right)\ ,
\nonumber\\
|\Psi_{4}({\bf r}_1,s_1;{\bf r}_2,s_2)\rangle&=&\frac{1}{\sqrt{2}}\,
\left[\phi_{\alpha}({\bf r}_1)\phi_{\beta}({\bf r}_2)-
\phi_{\beta}({\bf r}_1)\phi_{\alpha}({\bf r}_2)\right]
|\uparrow\rangle_1|\uparrow\rangle_2\ ,
\nonumber\\
|\Psi_{5}({\bf r}_1,s_1;{\bf r}_2,s_2)\rangle&=&\frac{1}{\sqrt{2}}\,
\left[\phi_{\alpha}({\bf r}_1)\phi_{\beta}({\bf r}_2)-
\phi_{\beta}({\bf r}_1)\phi_{\alpha}({\bf r}_2)\right]
|\downarrow\rangle_1|\downarrow\rangle_2\ ,
\nonumber\\
|\Psi_{6}({\bf r}_1,s_1;{\bf r}_2,s_2)\rangle&=&\frac{1}{2}
\left[\phi_{\alpha}({\bf r}_1)\phi_{\beta}({\bf r}_2)-
\phi_{\beta}({\bf r}_1)\phi_{\alpha}({\bf r}_2)\right]
\left(|\uparrow\rangle_1|\downarrow\rangle_2+
|\downarrow\rangle_1|\uparrow\rangle_2\right)\ .
\end{eqnarray}
Here, the six components of these two-electron sets are orthonormal, i.e.,
$\langle\Psi_{m}|\Psi_{n}\rangle=\delta_{m,n}$.
Moreover, using the above six states, the interacting Hamiltonian matrix can be cast into the form of

\begin{equation}
{\hat{\cal H}}=
\begin{pmatrix}
2\varepsilon_{\alpha}+U_{11} & U_{12} & U_{13} & 0 & 0 & 0\\
U_{21} & 2\varepsilon_{\beta}+U_{22} & U_{23} & 0 & 0 & 0\\
U_{31} & U_{32} & \varepsilon_{\alpha}+\varepsilon_{\beta}+U_{33} & 0 & 0 & 0\\
0 & 0 & 0 & \varepsilon_{\alpha}+\varepsilon_{\beta}+U_{44} & 0 & 0\\
0 & 0 & 0 & 0 & \varepsilon_{\alpha}+\varepsilon_{\beta}+U_{55} & 0\\
0 & 0 & 0 & 0 & 0 & \varepsilon_{\alpha}+\varepsilon_{\beta}+U_{66}
\end{pmatrix}\ ,
\label{add5}
\end{equation}
where $U_{mn}=\langle\Psi_{m}|{\cal U}_{\rm C}|\Psi_{n}\rangle=U^\ast_{nm}$ is the Coulomb matrix
element. Explicitly, we define the notations for the Coulomb matrix elements as

\begin{eqnarray}
{\cal M}_{\alpha\beta;\,\alpha^\prime\beta^\prime}&\equiv&\langle\phi_{\alpha},\phi_{\beta}|{\cal U}_{\rm C}
|\phi_{\alpha^\prime},\phi_{\beta^\prime}\rangle\ ,
\nonumber\\
{\cal M}_{\beta\alpha;\,\beta^\prime\alpha^\prime}&=&{\cal M}_{\alpha\beta;\,\alpha^\prime\beta^\prime}\ .
\label{add6}
\end{eqnarray}
Introducing the Fourier transform to the Coulomb potential, we are able to express its matrix
elements using the Coulomb and exchange integrals, i.e.,

\begin{equation}
\frac{1}{|{\bf r}_1-{\bf r}_2|}=\int\frac{d^2{\bf q}}{2\pi}\,
\frac{e^{i{\bf q}\cdot({\bf r}_1-{\bf r}_2)}}{|{\bf q}|}\ .
\end{equation}
This gives rise to

\begin{eqnarray}
U_{11}&=&{\cal M}_{\alpha\alpha;\,\alpha\alpha}
\equiv\langle\phi_{\alpha},\phi_\alpha|{\cal U}_{\rm C}
|\phi_\alpha,\phi_\alpha\rangle
\nonumber\\
&=&\frac{e^2}{2\epsilon_0\epsilon_{\rm r}}\int d^2{\bf q}\,
\frac{\left|{\cal F}_{\alpha\alpha}({\bf q})\right|^2}{|{\bf q}|}\ ,
\end{eqnarray}
where we have employed the form factor, given by

\begin{equation}
{\cal F}_{\alpha\beta}({\bf q})=\frac{1}{2\pi}\int d^2{\bf r}\,
\phi^\ast_{\alpha}({\bf r})\,e^{i{\bf q}\cdot{\bf r}}\,\phi_{\beta}({\bf r})
={\cal F}^\ast_{\beta\alpha}(-{\bf q})\ .
\end{equation}
Similarly,   we can obtain other nonzero matrix elements from

\begin{eqnarray}
U_{22}&=&{\cal M}_{\beta\beta;\,\beta\beta}=
\frac{e^2}{2\epsilon_0\epsilon_{\rm r}}\int \frac{d^2{\bf q}}{|{\bf q}|}
\left|{\cal F}_{\beta\beta}({\bf q})\right|^2\ ,
\nonumber\\
U_{12}&=&{\cal M}_{\alpha\alpha;\,\beta\beta}
=\frac{e^2}{2\epsilon_0\epsilon_{\rm r}}\int \frac{d^2{\bf q}}{|{\bf q}|}\,
{\cal F}_{\alpha\beta}({\bf q}){\cal F}_{\alpha\beta}(-{\bf q})\ ,
\nonumber\\
U_{13}&=&\frac{1}{\sqrt{2}}\left({\cal M}_{\alpha\alpha;\,\alpha\beta}
+{\cal M}_{\alpha\alpha;\,\beta\alpha}\right)
\nonumber\\
&=&\frac{e^2}{2\sqrt{2}\epsilon_0\epsilon_{\rm r}}\int \frac{d^2{\bf q}}{|{\bf q}|}
\left[{\cal F}_{\alpha\alpha}({\bf q}){\cal F}_{\alpha\beta}(-{\bf q})
+{\cal F}_{\alpha\beta}({\bf q}){\cal F}_{\alpha\alpha}(-{\bf q})\right]\ ,
\nonumber\\
U_{23}&=&\frac{1}{\sqrt{2}}\left({\cal M}_{\beta\beta;\,\alpha\beta}
+{\cal M}_{\beta\beta;\,\beta\alpha}\right)
\nonumber\\
&=&\frac{e^2}{2\sqrt{2}\epsilon_0\epsilon_{\rm r}}\int \frac{d^2{\bf q}}{|{\bf q}|}
\left[{\cal F}_{\beta\alpha}({\bf q}){\cal F}_{\beta\beta}(-{\bf q})
+{\cal F}_{\beta\beta}({\bf q}){\cal F}_{\beta\alpha}(-{\bf q})\right]\ ,
\nonumber\\
U_{33}&=&{\cal M}_{\alpha\beta;\,\alpha\beta}
+{\cal M}_{\alpha\beta;\,\beta\alpha}
=\frac{e^2}{2\epsilon_0\epsilon_{\rm r}}\int \frac{d^2{\bf q}}{|{\bf q}|}
\left[{\cal F}_{\alpha\alpha}({\bf q}){\cal F}_{\beta\beta}(-{\bf q})
+\left|{\cal F}_{\alpha\beta}({\bf q})\right|^2\right]\ ,
\nonumber\\
U_{44}&=&{\cal M}_{\alpha\beta;\,\alpha\beta}
-{\cal M}_{\alpha\beta;\,\beta\alpha}
=\frac{e^2}{2\epsilon_0\epsilon_{\rm r}}\int \frac{d^2{\bf q}}{|{\bf q}|}
\left[{\cal F}_{\alpha\alpha}({\bf q}){\cal F}_{\beta\beta}(-{\bf q})
-\left|{\cal F}_{\alpha\beta}({\bf q})\right|^2\right]
\nonumber\\
&=&U_{55}=U_{66}\ .
\end{eqnarray}
After we diagonalize the Hamiltonian matrix in Eq.\,(\ref{add5}), both the energy eigenvalues
and associated eigenstates can be obtained in a straightforward way.
\medskip
\par

Now, let us consider explicitly a harmonic confining  potential for  electrons within the $xy-$plane, i.e.,

\begin{equation}
{\cal V}_{\rm c}({\bf r})=\frac{1}{2}\,m^\ast\left(\omega^2_xx^2+\omega^2_yy^2\right)\ ,
\label{add7}
\end{equation}
with $\omega_y\ll\omega_x$. As a result, the orbital parts of the single-particle eigenstates can be written down as

\begin{equation}
\phi_{m,n}(x,\,y)=\psi_m(x)\,\psi_n(y)\ ,\ \ \ \ \varepsilon_{m,n}=\hbar\omega_x\left(m+1/2\right)+\hbar\omega_y\left(n+1/2\right)\ .
\label{add8}
\end{equation}
We will choose two eigenstates, $\phi_{\alpha}(x,\,y)=\psi_0(x)\,\psi_0(y)$
and $\phi_{\beta}(x,\,y)=\psi_1(x)\,\psi_0(y)$, where $\psi_n(x)$ is the one-dimensional oscillator
wavefunction for $n=0,\,1,\,2,\,\cdots$. This yields

\begin{eqnarray}
\phi_{\alpha}(x,\,y)&=&\left(\frac{1}{\pi\xi_x\xi_y}\right)^{1/2}
\exp\left(-\frac{x^2}{2\xi^2_x}\right)\exp\left(-\frac{y^2}{2\xi^2_y}\right)\ ,
\nonumber\\
{\cal F}_{\alpha\alpha}({\bf q})&=&\frac{1}{2\pi}
\exp\left(-\frac{q^2_x\xi^2_x}{4}\right)\exp\left(-\frac{q^2_y\xi^2_y}{4}\right)\ ,
\label{add9}
\end{eqnarray}
as well as

\begin{eqnarray}
\phi_{\beta}(x,\,y)&=&\left(\frac{1}{2\pi\xi_x\xi_y}\right)^{1/2}
\left(\frac{2x}{\xi_x}\right)\exp\left(-\frac{x^2}{2\xi^2_x}\right)\exp\left(-\frac{y^2}{2\xi^2_y}\right)\ ,
\nonumber\\
{\cal F}_{\beta\beta}({\bf q})&=&
\frac{1}{4\pi}\left(2-q^2_x\xi^2_x \right)
\exp\left(-\frac{q^2_x\xi^2_x}{4}\right)\exp\left(-\frac{q^2_y\xi^2_y}{4}\right)\ ,
\nonumber\\
{\cal F}_{\alpha\beta}({\bf q})&=&
\frac{iq_x\xi_x}{2\pi\sqrt{2}}
\exp\left(-\frac{q^2_x\xi^2_x}{4}\right)\exp\left(-\frac{q^2_y\xi^2_y}{4}\right)\ .
\label{add10}
\end{eqnarray}
Here, $\xi_x=\sqrt{\hbar/m^\ast\omega_x}$ and $\xi_y=\sqrt{\hbar/m^*\omega_y}$ are the confining lengths, and then,
the system dimension may be significantly larger along the $y$-direction compared to that in the $x$-direction.
\medskip
\par

We know from Eqs.\,(\ref{add8}) and (\ref{add9}) that eigenstates $\phi_{\alpha}({\bf r})$ and $\phi_{\beta}({\bf r})$ have opposite parity.
Consequently, we find ${\cal F}_{\alpha\alpha}({\bf q})={\cal F}_{\alpha\alpha}(-{\bf q})$, ${\cal F}_{\beta\beta}({\bf q})={\cal F}_{\beta\beta}(-{\bf q})$, and
${\cal F}_{\alpha\beta}({\bf q})+{\cal F}_{\alpha\beta}(-{\bf q})=0$. This directly leads to
$u_{13}=u_{23}=0$. Moreover, for a harmonic potential
the block part of the truncated Hamiltonian in Eq.\,(\ref{add5})  becomes

\begin{equation}
{\hat{\cal H}_{3\times3}}=
\begin{pmatrix}
2\varepsilon_{\alpha}+U_{11} & U_{12} & 0\\
U_{12} & 2\varepsilon_{\beta}+U_{22} & 0\\
0 & 0 & \varepsilon_{\alpha}+\varepsilon_{\beta}+U_{33}
\end{pmatrix}\ .
\label{add11}
\end{equation}
From this, we obtain two split energy eigenvalues for the states $\Psi_{1}$ and $\Psi_{2}$, given by

\begin{equation}
E_{1,2}\equiv E_{\pm}=\varepsilon_{\alpha}+\varepsilon_{\beta}
+\frac{1}{2}\left(U_{11}+U_{22}\right)\pm{\cal D}\ ,
\label{add12}
\end{equation}
and the uncoupled energy level $E_3=\varepsilon_{\alpha}+\varepsilon_{\beta}+U_{33}$ for
the state $\Psi_{3}$, as well as the triple-degenerate energy levels $E_4=E_5=E_6=\varepsilon_{\alpha}+\varepsilon_{\beta}+U_{44}$
for the states $\Psi_{4}$, $\Psi_{5}$ and $\Psi_{6}$.
In Eq.\,(\ref{add12}), the energy-level coupling ${\cal D}=\sqrt{[\varepsilon_{\beta}-\varepsilon_{\alpha}+(U_{22}-U_{11})/2]^2+|U_{12}|^2}$, and
the level splitting is $E_{+}-E_{-}=2{\cal D}>0$.
\medskip
\par

For evaluating $u_{11}$, $u_{22}$, $u_{33}$, $u_{12}$ and $u_{44}$, we require the following Coulomb integrals:

\[
{\cal I}_{\alpha\alpha;\,\alpha\alpha}=2\pi L_y\int \frac{d^2{\bf q}}
{|{\bf q}|}\,\left|{\cal F}_{\alpha\alpha}({\bf q})\right|^2
=\frac{L_y}{2\pi}\int_0^{2\pi} d\theta\,\int_0^\infty dq\, e^{-\alpha(\theta)q^2}
=\frac{L_y}{4\sqrt{\pi}}\int_0^{2\pi} \frac{d\theta}{\sqrt{\alpha(\theta)}}
\]
\begin{equation}
=\frac{L_y}{\sqrt{2\pi}}\int_0^\pi \frac{d\theta}{(\xi_x^2\cos^2\theta+\xi_y^2\sin^2\theta)^{1/2}}
=\frac{L_y}{\sqrt{2\pi}\xi_y}\int_0^\pi \frac{d\theta}{[1+(\gamma^2-1)\cos^2\theta]^{1/2}}\ ,
\label{add13}
\end{equation}
where we have used $q_x=q\cos\theta$, $q_y=q\sin\theta$,
$\theta$ is the angle between the wave vector ${\bf q}$ and $x$ axis, $\gamma\equiv\xi_x/\xi_y$,
and $\alpha(\theta)=(\xi_x^2\cos^2\theta+\xi_y^2\sin^2\theta)/2$. Additionally, we   obtain

\begin{eqnarray}
{\cal I}_{\beta\beta;\,\beta\beta}&=& 2\pi L_y\int \frac{d^2{\bf q}}
{|{\bf q}|}\,\left|{\cal F}_{\beta\beta}({\bf q})\right|^2=\frac{L_y}{8\pi}\int_0^{2\pi} d\theta\int_0^\infty dq\left(2-q^2\xi_x^2\cos^2\theta\right)^2\,e^{-\alpha(\theta)q^2}
\nonumber\\
&=&\frac{L_y}{4\pi}\int_0^\pi d\theta\left[I_1(\theta)+I_2(\theta)+I_3(\theta)\right]\ ,
\end{eqnarray}
where

\begin{eqnarray}
I_1(\theta)&=&4\int_0^\infty dq\,e^{-\alpha(\theta)q^2}=\frac{2\sqrt{\pi}}{\sqrt{\alpha(\theta)}}\ ,
\nonumber\\
I_2(\theta)&=& -4\xi_x^2\cos^2\theta\int_0^\infty dq\,q^2\,e^{-\alpha(\theta)q^2}
=-\frac{\sqrt{\pi}\xi_x^2\cos^2\theta}{\sqrt{\alpha^3(\theta)}}\ ,
\nonumber\\
I_3(\theta)&=&\xi_x^4\cos^4\theta\int_0^\infty dq\,q^4\,e^{-\alpha(\theta)q^2}
=\frac{3\sqrt{\pi}\xi_x^4\cos^4\theta}{8\sqrt{\alpha^5(\theta)}}\ .
\end{eqnarray}
By combining the results for $I_1(\theta)$, $I_2(\theta)$ and $I_3(\theta)$, this leads to

\begin{eqnarray}
&& {\cal I}_{\beta\beta;\,\beta\beta}=\frac{L_y}{\sqrt{2\pi}}\int_0^\pi \frac{d\theta}{(\xi_x^2\cos^2\theta+\xi_y^2\sin^2\theta)^{1/2}}
\nonumber\\
&\times&  \left[1-\frac{\xi_x^2\cos^2\theta}{\xi_x^2\cos^2\theta+\xi_y^2\sin^2\theta}
+\frac{3\xi_x^4\cos^4\theta}{4(\xi_x^2\cos^2\theta+\xi_y^2\sin^2\theta)^2}\right]
=\frac{L_y}{\sqrt{2\pi}\xi_y}\int_0^\pi \frac{d\theta}{[1+(\gamma^2-1)\cos^2\theta]^{1/2}}
\nonumber\\
&\times&\left[1-\frac{\gamma^2\cos^2\theta}{1+(\gamma^2-1)\cos^2\theta}
+\frac{3\gamma^4\cos^4\theta}{4[1+(\gamma^2-1)\cos^2\theta]^2}\right]\ .
\label{add14}
\end{eqnarray}
In a similar way, we find

\begin{eqnarray}
{\cal I}_{\alpha\beta;\,\alpha\beta}&=&2\pi L_y\int \frac{d^2{\bf q}}
{|{\bf q}|}\,{\cal F}_{\alpha\alpha}({\bf q}){\cal F}_{\beta\beta}(-{\bf q})=\frac{L_y}{4\pi}\int_0^{2\pi} d\theta\int_0^\infty dq\left(2-q^2\xi_x^2\cos^2\theta\right)\,e^{-\alpha(\theta)q^2}
\nonumber\\
&=& \frac{L_y}{2\pi}\int_0^\pi d\theta\left[J_1(\theta)+J_2(\theta)\right]\ ,
\end{eqnarray}
where

\begin{eqnarray}
J_1(\theta)&=&2\int_0^\infty dq\,e^{-\alpha(\theta)q^2}=\frac{\sqrt{\pi}}{\sqrt{\alpha(\theta)}}\ ,
\nonumber\\
J_2(\theta)&=&-\xi_x^2\cos^2\theta\int_0^\infty dq\,q^2\,e^{-\alpha(\theta)q^2}
=-\frac{\sqrt{\pi}\xi_x^2\cos^2\theta}{4\sqrt{\alpha^3(\theta)}}\ .
\end{eqnarray}
By combining these results for $J_1(\theta)$ and $J_2(\theta)$, we  have

\begin{eqnarray}
{\cal I}_{\alpha\beta;\,\alpha\beta}&=&\frac{L_y}{\sqrt{2\pi}}\int_0^\pi
\frac{d\theta}{(\xi_x^2\cos^2\theta+\xi_y^2\sin^2\theta)^{1/2}}\left[1-
\frac{\xi_x^2\cos^2\theta}{2(\xi_x^2\cos^2\theta+\xi_y^2\sin^2\theta)}\right]
\nonumber\\
&=& \frac{L_y}{\sqrt{2\pi}\xi_y}\int_0^\pi \frac{d\theta}{[1+(\gamma^2-1)
\cos^2\theta]^{1/2}}\left[1-\frac{\gamma^2\cos^2\theta}{2[1+(\gamma^2-1)\cos^2\theta]}\right]
\nonumber\\
&=& \frac{L_y}{2\sqrt{2\pi}\xi_y}\int_0^\pi d\theta\,\frac{2+(\gamma^2-1)\cos^2\theta}{[1+(\gamma^2-1)\cos^2\theta]^{3/2}}\ .
\label{add15}
\end{eqnarray}
The last integral is calculated as

\begin{eqnarray}
{\cal I}_{\alpha\beta;\,\beta\alpha}&=& I_{\alpha\alpha;\,\beta\beta}=2\pi L_y\int \frac{d^2{\bf q}}
{|{\bf q}|}|\,{\cal F}_{\alpha\beta}({\bf q}){\cal F}_{\alpha\beta}(-{\bf q})=
\frac{L_y}{4\pi}\int_0^{2\pi} d\theta\int_0^\infty dq\,q^2\xi_x^2\cos^2\theta\,e^{-\alpha(\theta)q^2}
\nonumber\\
&=& \frac{L_y}{8\sqrt{\pi}}\int_0^\pi d\theta\,\frac{\xi_x^2\cos^2\theta}{\sqrt{\alpha^3(\theta)}
}=\frac{1}{2\sqrt{2\pi}}\int_0^\pi d\theta\,\frac{\xi_x^2\cos^2\theta}{(\xi_x^2\cos^2\theta+\xi_y^2\sin^2\theta)^{3/2}}
\nonumber\\
&=&  \frac{L_y}{2\sqrt{2\pi}\xi_y}\int_0^\pi d\theta\,\frac{\gamma^2\cos^2\theta}{[1+(\gamma^2-1)\cos^2\theta]^{3/2}}\ .
\label{add16}
\end{eqnarray}
\medskip
\par

It is important to note that we have assumed a quasi-continuum energy spectrum for a traveling
quasiparticle in the longitudinal direction, in contrast with split energy levels in the
transverse direction. In this case, the pair of electrons forming the quasiparticle
always have the lowest transverse energy  plus a free electron-like kinetic energy,
resulting from the longitudinal motion. However,
the Coulomb interaction between a pair of electrons in this cluster
will significantly modify the ``subband edges'' ($E^{(p)}_{j}$) due to
quantization in the transverse direction.
\medskip
\par

All quasiparticles, except the transported one, may be treated as a ``background''
making up the total electron density and have a quasiparticle chemical potential
$\mu_p$ which is determined using Eq.\,(\ref{e2}).
For the interacting two-electron states, by using the above derivations,
their energy levels  $E_{j}^{(p)}=E_{j,k_y=0}^{(p)}$ are calculated as $E_{1}^{(p)}\equiv E_{-}^{(p)}=\varepsilon_0+\varepsilon_1+
 (u_{11}+u_{22})/2-\Delta_{\rm C}$, $E_{2}^{(p)}\equiv E_{+}^{(p)}=\varepsilon_0+\varepsilon_1+
(u_{11}+u_{22})/2+\Delta_{\rm C}$, $E_{3}^{(p)} =\varepsilon_0+\varepsilon_1+u_{33}$ and
$E_{4}^{(p)} =\varepsilon_0+\varepsilon_1+u_{44}$, where that Coulomb coupling term for the two-electron
anticrossing states is given by $\Delta_{\rm C}=\sqrt{\left[\varepsilon_{1}-\varepsilon_{0}+(u_{22}-u_{11})/2\right]^2+|u_{12}|^2}$.
In this notation, the
single-particle energy levels for the harmonic-potential model with anisotropic harmonic frequencies $\omega_x$
and $\omega_y$ in the transverse ($x$) and longitudinal ($y$) directions, respectively,
are: $\varepsilon_0=(\hbar\omega_x+\hbar\omega_y)/2$ and
$\varepsilon_1=(3\hbar\omega_x+\hbar\omega_y)/2$, while the
introduced Coulomb interaction energies are found to be
$u_{11}/E_{\rm c}=N_0^2\,{\cal I}_{00,00}$, $u_{12}/E_{\rm c}=N_0N_1\,{\cal I}_{00,11}$,
$u_{22}/E_{\rm c}=N_1^2\,{\cal I}_{11,11}$, $u_{33}/E_{\rm c}=N_0N_1\left({\cal I}_{01,01}+{\cal I}_{01,10} \right)$
and $u_{44}/E_{\rm c}=N_0N_1\left(3\,{\cal I}_{01,01}-{\cal I}_{01,10}\right)$, where
$E_{\rm c}=e^2/4\pi\epsilon_0\epsilon_rL_y$ in terms of the channel length $L_y$
and the background dielectric constant $\epsilon_r$, $N_n=\{\exp[(\varepsilon_n-\mu_0)/k_{\rm B}T]+1\}^{-1}$
($n=0,\,1$) is the single-particle level occupation factor, and
$\mu_0(T,\,n_{\rm 1D})$ is the single-electron chemical potential within the channel.
At the time when a quasiparticle enters a conduction channel, it will
occupy single-particle energy levels $\varepsilon_0$ and $\varepsilon_1$,
i.e., occupying the same one or different levels. Such a selection is determined
from the subband occupation by the sea of electrons within the channel.
After these two noninteracting electrons are injected into the channel, they will
interact with each other through either the intrasubband or intersubband Coulomb coupling.
The ballistic injection of two noninteracting electrons and the existence of a sea of
electrons  in the conduction channel are reflected through the inclusion of these two
level occupation factors. The Coulomb integral is represented by ${\cal I}_{\alpha\beta,\alpha'\beta'}(\gamma)$ for
$\alpha,\beta,\alpha',\beta'=0,1$ if we only consider interacting electron states formed from
the lowest (`$0$') and first excited (`$1$') single-particle states.
\medskip
\par

The channel width $W_x$ and length $L_y$ are directly related to the frequencies $\omega_x$ and $\omega_y$
of the 2D harmonic-confining potential by $W_x=\sqrt{4\hbar/m^\ast\omega_x}$ and
$L_y=\sqrt{4\hbar/m^\ast\omega_y}$, respectively. Therefore, we get the simple relations, i.e.,
$\xi_x=\sqrt{\hbar/m^\ast\omega_x}=W_x/2$, $\xi_y=\sqrt{\hbar/m^\ast\omega_y}=L_y/2$, and
and $\gamma\equiv\xi_x/\xi_y=W_x/L_y\equiv{\cal R}$.
Furthermore, for ${\cal R}\gg 1$, we find that ${\cal I}_{\alpha\beta,\alpha',\beta'}({\cal R})$ scales as $1/{\cal R}$
for $\alpha,\beta,\alpha',\beta'=0,1$ .
\medskip
\par

In the random-phase approximation (RPA), the static dielectric function at
low temperature for screening for an electron density
$n_{\rm 1D}$ and $\epsilon_r$ in the channel, may be expressed as\,\cite{lyo1}

\begin{equation}
\epsilon_{\rm 1D}(q)=1-\left(\frac{m^\ast e^2}{2\pi\epsilon_0\epsilon_r\hbar^2n_{\rm 1D}}\right)
\ln\left(\frac{|q|W_x}{2}\right)\ ,
\label{add18}
\end{equation}
where the wave vector $q\sim k_{\rm F}=\pi n_{\rm 1D}/2$. For the parameters chosen in
our  numerical calculations, we found that the effect due to static screening
may be neglected. On the other hand, the static dielectric function for shielding by surface
gate electrodes of the electron-electron interaction may be modeled as $\epsilon_{\rm G}(q)=1+\coth(qd)$,
for which we may take
$q\sim k_{\rm F}=\pi n_{\rm 1D}/2$ and $d$ represents the gate insulator thickness\,\cite{azin}.
For the parameters used in our numerical calculations, we found that  shielding
of the interaction between two-electron states may also be neglected.
\medskip
\par

From the calculated $J^{(\alpha=0)}(V_b,\,T,\,n_{\rm 1D})$ in Eq.\,(\ref{e1}),
the electrical conductance $G(T,\,n_{\rm 1D})$ for interacting two-electrons
may be expressed as\,\cite{lyo5}

\begin{equation}
G(T,\,n_{\rm 1D})= \frac{J^{(\alpha=0)}(V_b,\,T,\,n_{\rm 1D})}{V_b}\ .
\label{e4}
\end{equation}
We  now present  our numerical results and
their relationship to  recently reported experimental data in Ref.\,[\onlinecite{pepper}].

\section{Numerical Results and Experimental Data}
\label{sec:3}

In our numerical calculations, we use the following parameters: $T=10\,$mK, $V_b=0.01\,$mV, $L_y=400\,$nm,
$\epsilon_r=12$, $m^\ast=0.067\,$m$_0$ ($m_0$ is the  free-electron mass). The chosen ${\cal R}$ values are
indicated in the figure captions. Specifically, we denote the quantum ballistic transport of two-electron
states with anticrossing levels through a conduction channel as one moving through either one
of two states $E^{(p)}_{\pm}$.
\medskip
\par

For clarity, we point out that as two electrons are injected into a conduction channel, they may
occupy specific single-particle subbands for their ballistic transport. The selection rule
is determined by the occupation factor of the electrons already sustained within the conduction
channel. During the time interval that the two injected moving electrons remain within the
channel, they may interact with each other through either the intrasubband or the
 intersubband Coulomb coupling. We emphasize that the linear density  for confined electrons
within the channel may be kept constant even when the channel width is varied. However, for this to occur,
the Fermi energy must automatically adjust itself to accommodate all electrons and additional
subbands will be populated with reduced energy level separations. Specifically, although the Fermi
energy is reduced, the number of electrons in the channel is not changed at all.
Furthermore, even when the Fermi energy is reduced, the second level can still be populated due
to reduced level separation at the same time so as to keep the number of electrons in the channel a constant.
Clearly, enhancement of the  Coulomb interaction is not solely determined by the electron density,
since it also depends on how electrons are distributed. For the Coulomb effect on the two-electron states,
the inclusion of a new populated two-electron state, where one electron stays in a lower-energy level
while the other electron populates a higher level, will introduce a new Coulomb-interaction channel for
the two-electron states.

\subsection{Two-Electron Energies within the Channel}
\label{sec:3-1}

\begin{figure}[t]
\centering
\includegraphics[width=0.65\textwidth]{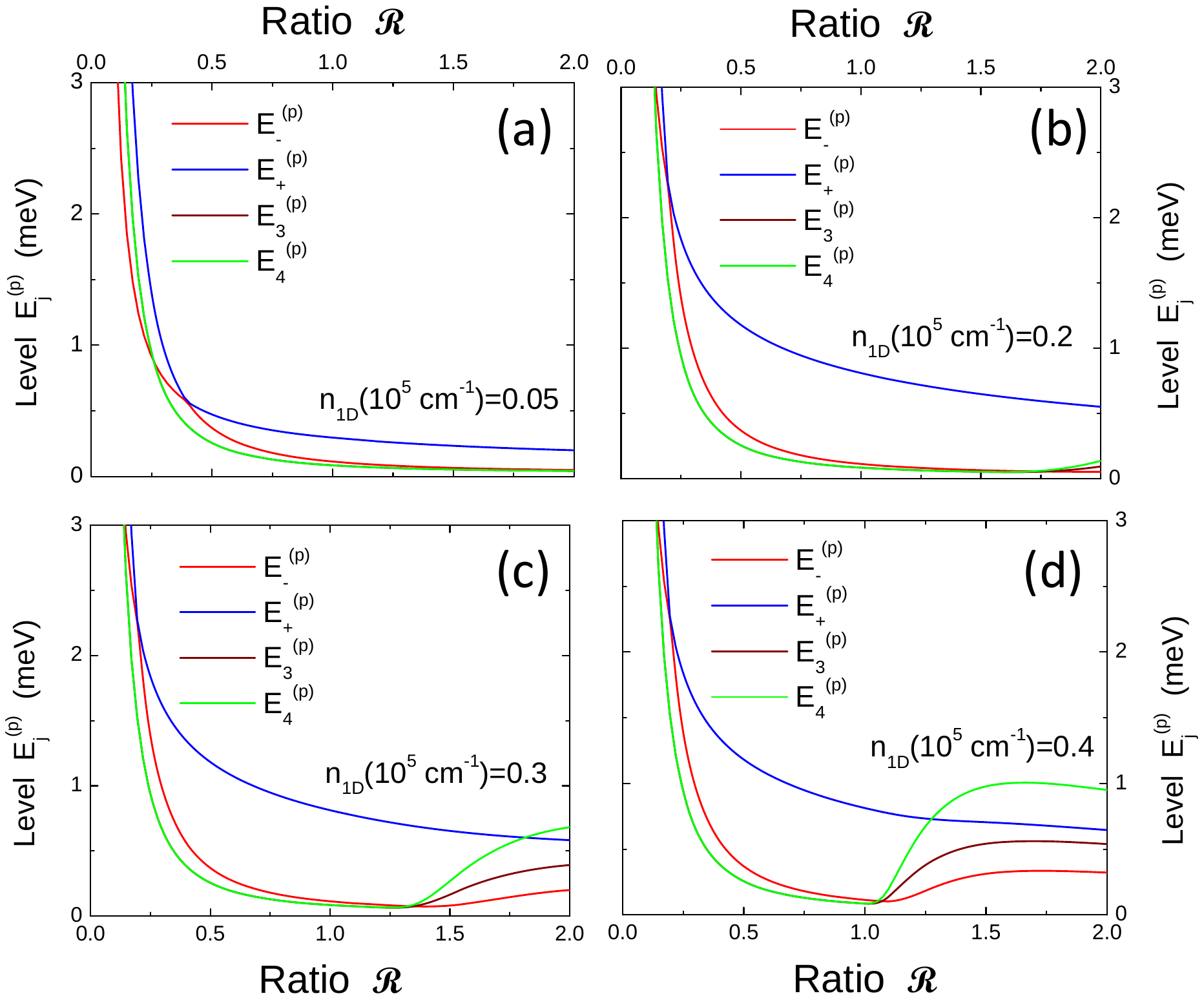}
\caption{(Color online) Plots of cluster energy levels $E_j^{(p)}$ as a function of geometry ratio ${\cal R}$ with four different values of linear electron density $n_{\rm 1D}$.}
\label{FIG:5}
\end{figure}

We know that as the transverse confinement becomes weaker (or the ${\cal R}$ value is increased), the
kinetic part of the energy levels $E_j^{(p)}$ of a two-electron state will decrease like as $1/{\cal R}^2$
for fixed $L_y$. On the other hand, the Coulomb interaction only scales as $1/{\cal R}$ as per our discussion
preceding Eq.\,(\ref{add18}). Consequently, the significance of the Coulomb interaction is expected to
increase relatively by increasing ${\cal R}$. Moreover, the level separation will be reduced by increasing
${\cal R}$, leading to occupation of the second energy level for fixed electron density. Therefore, the
additional Coulomb repulsion between two electrons on different single-particle energy levels must be considered.
This effect can be seen from Figs.\,\ref{FIG:5}(b), \ref{FIG:5}(c) and \ref{FIG:5}(d) as the upward shifting
of energy levels $E^{(p)}_-$ and $E^{(p)}_3$  (as $N_1>0$) in the region of ${\cal R}>1$ as
$n_{\rm 1D}\geq 0.2\times 10^{5}$\,cm$^{-1}$. At the same time, the $E^{(p)}_4$ level is pushed upward
above the $E^{(p)}_3$ level due to the enhanced Coulomb interaction for ${\cal R}>1$.
On the other hand, for the $E^{(p)}_+$ two-electron state, which is associated with two excited-state electrons,
it is largely dominated by the kinetic energy part for the whole range of ${\cal R}$ shown in this figure.
When $n_{\rm 1D}$ is further increased, the Coulomb repulsion effect pushes into the intermediate
confinement regime (${\cal R}\sim 1$) in Fig.\,\ref{FIG:5}(d). Due to the combined effect  of these two factors,
we observe the recovery of the ground state $E^{(p)}_-$ level in Fig.\ \ref{FIG:5}(d) for large values of
${\cal R}$ and $n_{\rm 1D}$ (where the Coulomb energy is dominant) from that in Fig.\ \ref{FIG:5}(a) for
small values of ${\cal R}$ and $n_{\rm 1D}$ (where the kinetic energy of electrons is dominant).
It is interesting to see that the Coulomb interaction between electrons stands out to give rise
to a pushing up of three energy levels and the recovery of the the ground $E^{(p)}_-$ level at the same time
in an intermediate confinement regime (${\cal R}\gtrsim 1$) between the strong (scaling as fast
drop $1/{\cal R}^2$ for ${\cal R}<1$)  and weak (scaling as slow drop $1/{\cal R}$ for ${\cal R}\gg 1$)
confinement regimes.

\subsection{Ballistic Conductance within a quasi-1D Channel}
\label{sec:3-2}

\begin{figure}[t]
\centering
\includegraphics[width=0.65\textwidth]{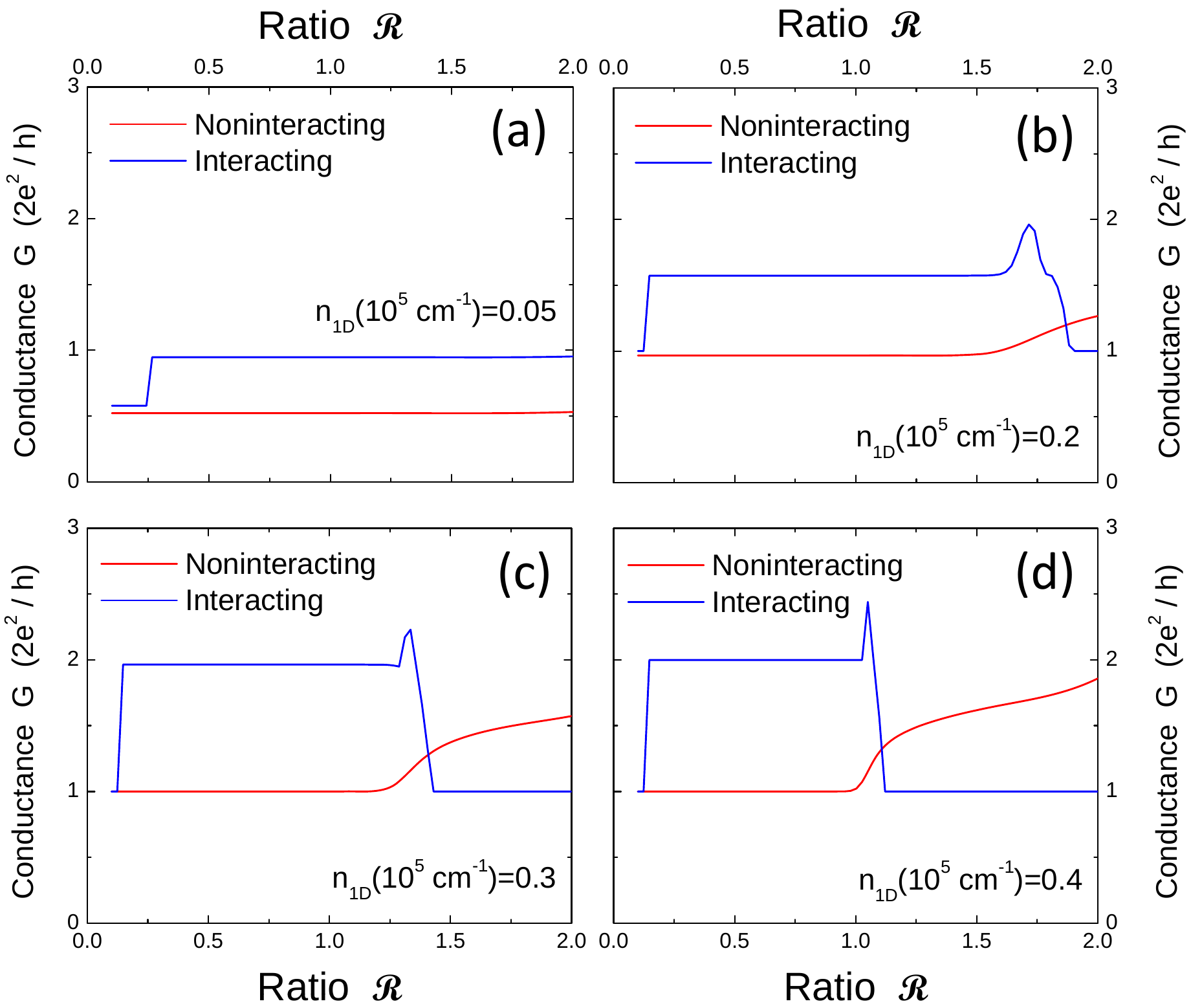}
\caption{(Color online) Plots of conductance $G$ as a function of geometry ratio ${\cal R}$ with four different values of linear electron density $n_{\rm 1D}$ for both noninteracting and interacting cases.}
\label{FIG:6}
\end{figure}

The recovery of the ground-state in Fig.\,\ref{FIG:5} plays a significant role on both the distribution of
conductance plateaus and the interplay from interaction effects, as displayed in Fig.\,\ref{FIG:6}.
We know the Coulomb coupling may be neglected for small $n_{\rm 1D}$, where
the $2e^2/h$ conductance plateau is found for the interacting two-electron state as shown in \ref{FIG:6}(a) with
almost all values of ${\cal R}$. As $n_{\rm 1D}$ increases to $0.2\times 10^{5}$\,cm$^{-1}$ in \ref{FIG:6}(b),
the $2e^2/h$ plateau shown in \ref{FIG:6}(a) disappears except for its reappearance very close to ${\cal R}=2.0$.
If the value of $n_{\rm 1D}$ gets even larger, as seen from Figs.\ \ref{FIG:6}(c) and \ref{FIG:6}(d),
the new $4e^2/h$ conductance plateau occurs for an interacting two-electron state, corresponding to
the population of the degenerate lowest $E^{(p)}_3$ and $E^{(p)}_4$ energy
levels after their level crossing with another $E^{(p)}_-$ state.
However, when ${\cal R}$ further increases above one in the very-weak confinement regime,
the ground-state recovery, as discussed in Figs.\,\ref{FIG:5}(c) and \ref{FIG:5}(d), enforces
the reoccurrence of the $2e^2/h$ conductance plateau due to the Coulomb
repulsion between electrons in the central region of the channel.

\subsection{Dependence of interacting electron energy on Linear Density}
\label{sec:3-3}

\begin{figure}[t]
\centering
\includegraphics[width=0.65\textwidth]{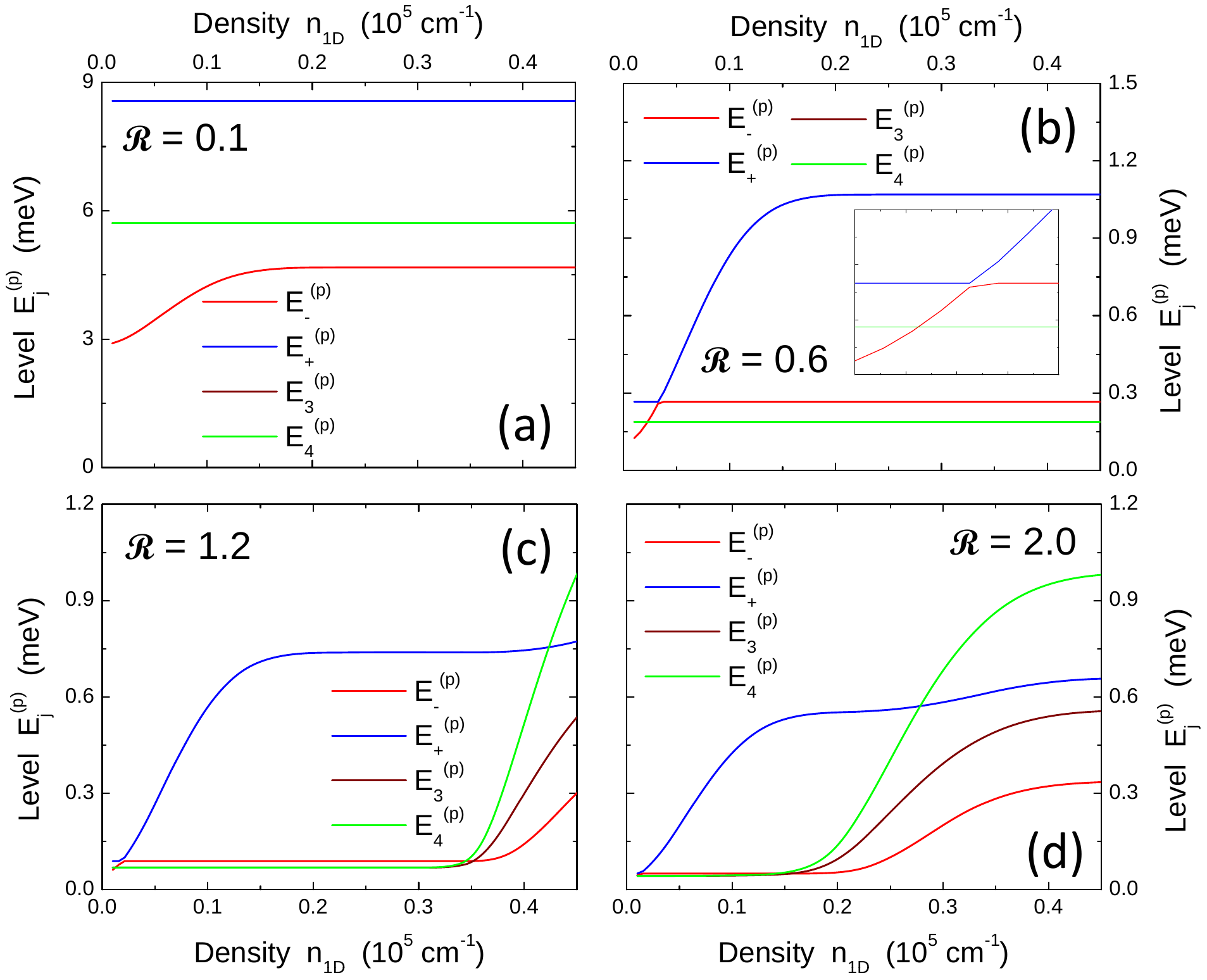}
\caption{(Color online) Plots of cluster energy levels $E_j^{(p)}$ as a function of linear electron density $n_{\rm 1D}$ with four different values of geometry ratio ${\cal R}=W_x/L_y$.
Inset in \ref{FIG:2}(b) brings us a blow-out view for the anticrossing of energy levels $E_{-}^{(p)}$ and $E_{+}^{(p)}$.}
\label{FIG:2}
\end{figure}

When electrons interact with each other, their energy levels $E_j^{(p)}$ are expected
to depend on the electron density $n_{\rm 1D}$, as shown in Fig.\,\ref{FIG:2}.
When the geometry ratio ${\cal R}=W_x/L_y$ is small for strong confinement in Fig. \ref{FIG:2}(a),
only the ground state $E^{(p)}_{-}$ is affected by varying $n_{\rm 1D}$ due  primarily to $N_0\neq 0$
in this case. As ${\cal R}$ is increased to $0.6$ in Fig.\ \ref{FIG:2}(b), both the level crossing between
$E^{(p)}_{-}$ of the anticrossing state with the degenerate state $E^{(p)}_3=E^{(p)}_4$ and the level anticrossing
between $E^{(p)}_{-}$ and $E^{(p)}_{+}$ states occur at lower densities.
As ${\cal R}>1$, the Coulomb interaction between electrons becomes much stronger, as presented in Figs.\
 \ref{FIG:2}(c)  and \ref{FIG:2}(d). Therefore, both $E^{(p)}_3$ and $E^{(p)}_4$ levels are pushed
up significantly at higher densities (i.e., $N_1>0$), leading to a recovery of the ground
state to $E^{(p)}_{-}$. At the same time, the $E^{(p)}_4$ state in Figs.\ref{FIG:2}(c)
and \ref{FIG:2}(d) changes from the degenerate ground state at lower $n_{\rm 1D}$ to the highest-energy
state at higher $n_{\rm 1D}$.  Furthermore, under a transverse magnetic field, we expect that
the $E^{(p)}_3$ state should decouple from the magnetic field due to total spin $S=0$, while the
degenerate $E^{(p)}_4$ state with total spin $S=1$ will be split into three by the Zeeman effect,
leading to new $e^2/h$ and $3e^2/h$ conductance plateaus\,\cite{half}.

\subsection{Conductance for two interacting and noninteracting Electron Pairs}
\label{sec:3-4}

\begin{figure}[t]
\centering
\includegraphics[width=0.65\textwidth]{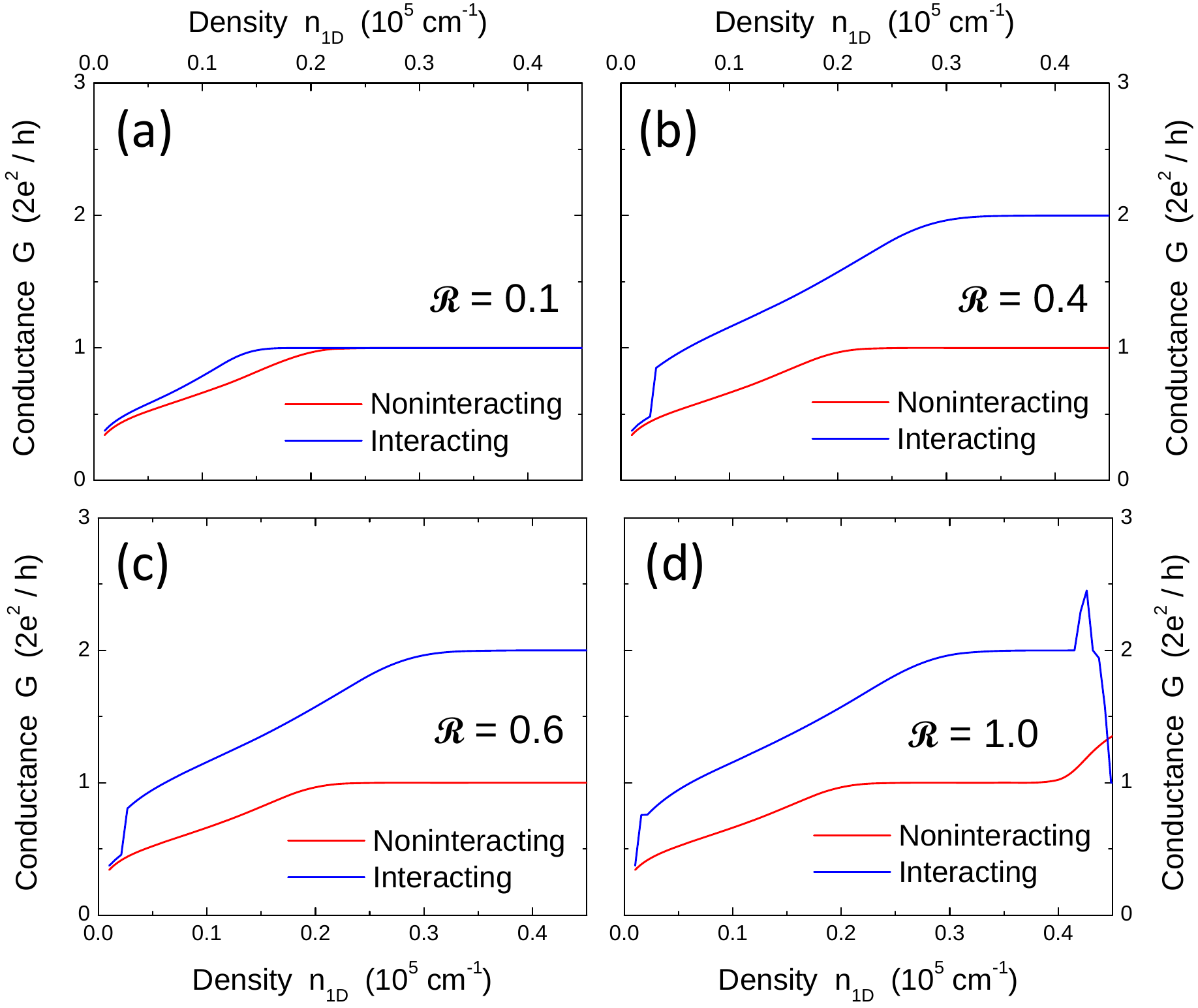}
\caption{(Color online) Plots of conductance $G$ as a function of linear electron density $n_{\rm 1D}$ with four different values of geometry ratio ${\cal R}$ for both noninteracting and interacting cases.}
\label{FIG:3}
\end{figure}

Figure\ \ref{FIG:3} presents a comparison of the conductance $G$ for both a noninteracting and
interacting two-electron state in the range of $0.1\leq {\cal R}\leq 1$. For very strong confinement
in \ref{FIG:2}(a), the Coulomb-interaction effect becomes negligible in comparison with the dominant
kinetic energy of electrons and a conductance $2e^2/h$ plateau remains with increasing $n_{\rm 1D}$.
On the other hand, as ${\cal R}$ goes up to $0.4$ in \ref{FIG:3}(b) and $0.6$ in \ref{FIG:3}(c) for
cases with strong confinement,  although $G$ for a noninteracting two-electron state remains
largely unchanged, for an interacting two-electron state, the conductance $2e^2/h$ plateau in
Fig.\ \ref{FIG:3}(a) is completely destroyed by the Coulomb interaction and replaced by a new
$4e^2/h$ plateau.  This unique feature is attributed to the result of both a level-crossing
and a level anticrossing observed in Fig.\,\ref{FIG:2}(b). However, the new $4e^2/h$ conductance plateau
is severely perturbed at higher densities by a sharp spike and a followed  by a deep dip to the lower
$2e^2/h$ plateau as ${\cal R}=1$ for intermediate confinement in \ref{FIG:3}(d).
\medskip
\par

Although the dimensionless Coulomb integrals do not depend on the linear electron density
 $n_{\rm 1D}$, the energy levels $E_j^{(p)}\sim\{u_{ij}\}$ for a two-electron cluster
is proportional to the occupation factors ($N_0$ and $N_1$) in addition to these Coulomb
integrals. Moreover, these occupation factors are determined by the chemical potential
$\mu_n$ for noninteracting electrons through the Fermi function for fixed $n_{\rm 1D}$.
On the other hand, the cluster chemical potential $\mu_p$, determined by Eq.\,(\ref{e2}),
controls the behavior of cluster ballistic transport in the presence of a bias voltage $V_b$.

\subsection{Conductance for weak Confinement}
\label{sec:3-5}

\begin{figure}[t]
\centering
\includegraphics[width=0.65\textwidth]{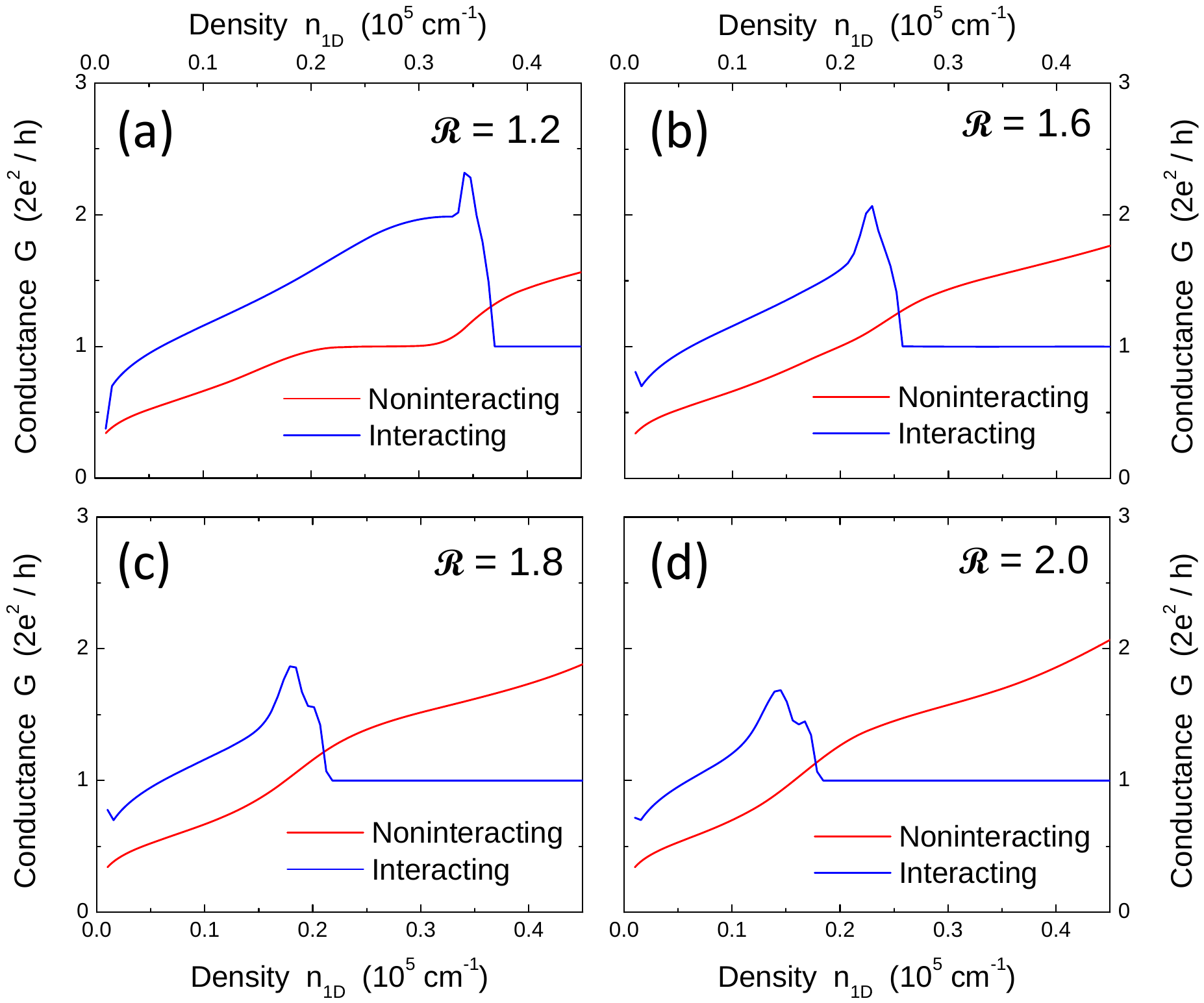}
\caption{(Color online) Plots of conductance $G$ as a function of linear electron density $n_{\rm 1D}$ with another four different values of geometry ratio ${\cal R}$ for both noninteracting and interacting cases.}
\label{FIG:4}
\end{figure}

We present in Fig.\,\ref{FIG:4} the change in the conductance plateau with increasing ${\cal R}$ in the
weak confinement regime.  When ${\cal R}\geq 1.6$, conductance plateaus for the noninteracting
two-electron state are washed out in Figs.\  \ref{FIG:4}(b)- \ref{FIG:4}(d) due to very small
single-particle energy level separation compared to the thermal energy $k_{\rm B}T$.
It is also evident that the incomplete $4e^2/h$ conductance plateau in Fig.\  \ref{FIG:4}(a)
for the interacting two-electron state is completely destroyed in this regime.
However, the recovery of the single-particle-like $2e^2/h$  plateau, as displayed in
Fig.\,\ref{FIG:3}(a), is found in Fig.\,\ref{FIG:4}. Additionally, the $2e^2/h$ plateau further
expands and extends to lower and lower electron densities as ${\cal R}$ increases to $2.0$
in Fig.\,\ref{FIG:4}(d). This unique reoccurrence feature can be fully accounted for  by the
rising energy levels at higher densities due to the relatively enhanced Coulomb repulsion
as shown in Figs.\,\ref{FIG:2}(c) and \ref{FIG:2}(d).
\medskip
\par

As displayed in Fig.\,\ref{FIG:2}, both $E_3^{(p)}$ and $E_4^{(p)}$ remain  degenerate for all chosen values of
$n_{\rm 1D}$ as far as ${\cal R}<1$ or alternatively for only small $n_{\rm 1D}$ values as ${\cal R}>1.2$.
The level-crossing between $E_-^{(p)}$ and the degenerate levels $E_3^{(p)}$ and $E_4^{(p)}$  is the reason behind
the upward jump of the conductance from $2e^2/h$ to $4e^2/h$, as can be seen from Fig.\,\ref{FIG:3}.
However, the degeneracy of the
$E_3^{(p)}$ and $E_4^{(p)}$ levels may be lifted by an enhanced Coulomb repulsion for ${\cal R}>1$ as well as for
large values of $n_{\rm 1D}$. Consequently, the subsequent downward dip in the conductance from
$4e^2/h$ to $2e^2/h$ is observed in Fig.\,\ref{FIG:4}.
\medskip
\par

\begin{figure}[t]
\centering
\includegraphics[width=0.75\textwidth]{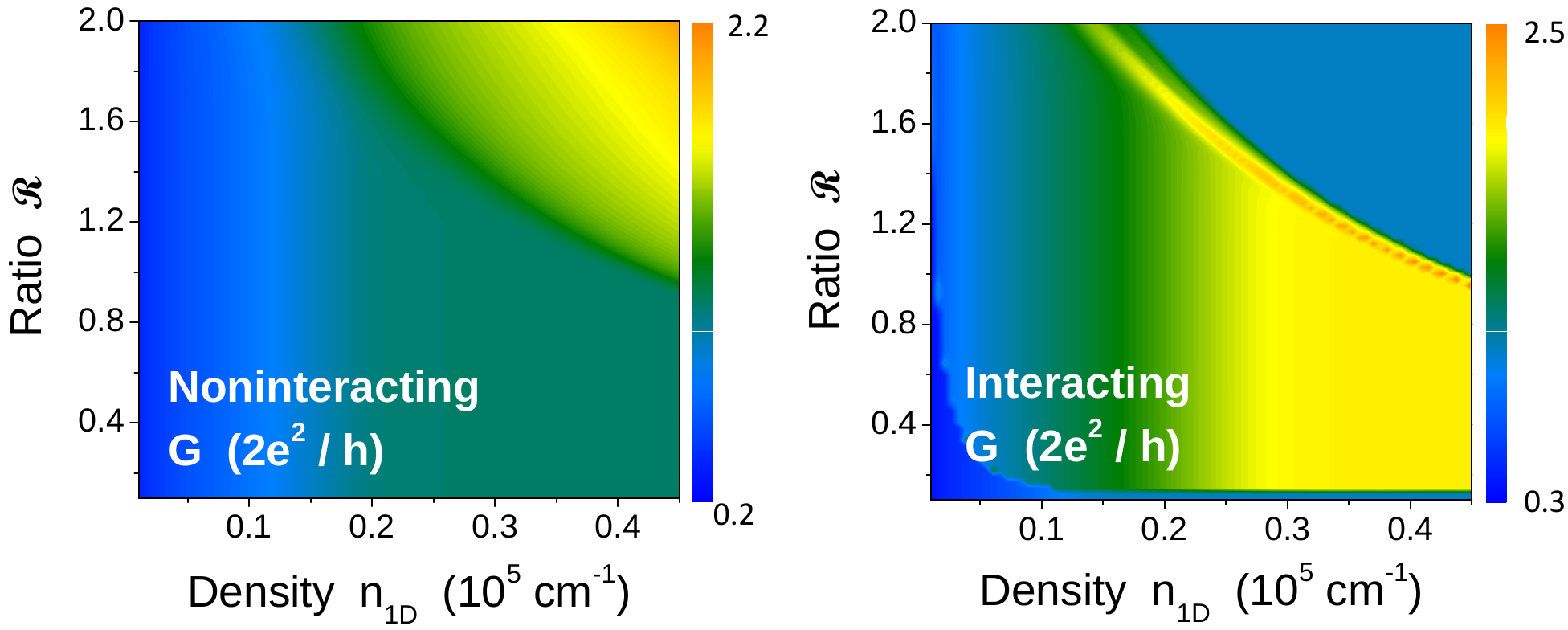}
\caption{(Color online) Contour plots of conductance $G$ as functions of both linear electron density $n_{\rm 1D}$ and geometry ratio ${\cal R}$ for either noninteracting (left panel) or interacting (right panel) case.
As labeled by the color bars in this figure, the color scales (from blue up to orange) are $[0.2,\,2.2]$ (left) and $[0.3,\,2.5]$ (right), respectively.}
\label{FIG:7}
\end{figure}

In order to acquire a complete picture of the quantum ballistic transport of interacting two-electron
states passing through a quasi-1D conduction channel, we present the contour plots of electron
conductance $G$ as functions of ${\cal R}$ and $n_{\rm 1D}$ in Fig.\,\ref{FIG:7} for both noninteracting
and interacting two-electron states as a direct comparison. By comparing Fig.\,\ref{FIG:7}(a) with
Fig.\,\ref{FIG:7}(b), we find that the effect of the Coulomb coupling becomes most dominant in the upper
right-hand  corner of Fig.\,\ref{FIG:7}(b) within a weak confinement regime and with a relatively high
electron density at the same time.  In this case,  a gradually increasing conductance for noninteracting
electrons is replaced by a $2e^2/h$ conductance plateau. This is due to the Coulomb repulsion
in interacting two-electron states.  In addition, another $4e^2/h$ conductance plateau shows up in
the lower right-hand corner of Fig.\,\ref{FIG:7}(b).   This is separated by a spike in $G$ from
the upper right-hand corner. In this  region,  confinement is intermediate or strong but the
electron density is high.
\medskip
\par

\begin{figure}[t]
\centering
\includegraphics[width=0.75\textwidth]{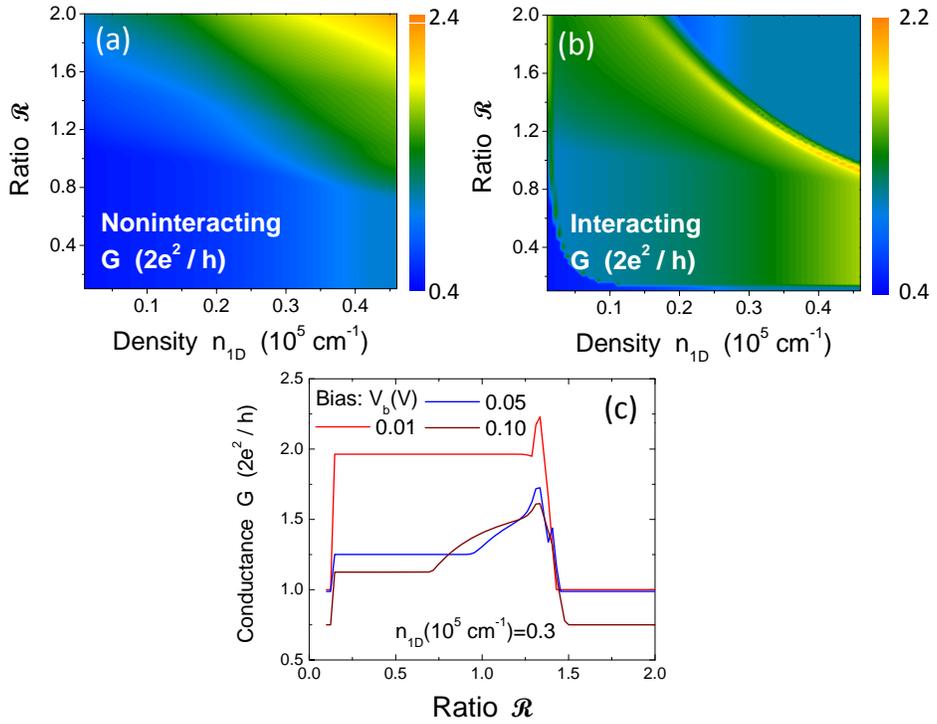}
\caption{(Color online) Contour plots of $G$ as functions of both linear electron density $n_{\rm 1D}$ and geometry ratio ${\cal R}$ for either noninteracting [(a)] or interacting [(b)] case at $V_b=0.05\,$V, as well as the
plot of $G$ as a function of ${\cal R}$ [(c)] at $n_{\rm 1D}=0.3\times 10^5\,$cm$^{-2}$ for three different values of bias voltage $V_b$.}
\label{FIG:8}
\end{figure}

In our numerical results presented above, we limit the bias voltage $V_b$ to a very small value ($0.01\,$mV),
where $G$ becomes essentially independent of $V_b$. The increase of $V_b$ can induce a ``hot-carrier''
effect and reduce the ballistic conductance with increasing temperature, as presented in Fig.\,\ref{FIG:8},
where the conductances $G$ for both  noninteracting [in Fig.\,\ref{FIG:8}(a)] and interacting
[in Fig.\,\ref{FIG:8}(b)]  with $V_b=0.05\,$mV are compared with each other. From Figs.\,\ref{FIG:7}(a)
and \ref{FIG:7}(b),  we find $G$ for noninteracting electrons has been changed qualitatively for
different values of $V_b$, although $G$ for electron clusters is only modified quantitatively.
We further demonstrate such a bias dependent effect on $G$ of electron clusters in Fig.\,\ref{FIG:8}(c),
where three different values of $V_b$ are chosen for $n_{\rm 1D}=0.3\times 10^5\,$cm$^{-1}$. As can be seen
from Fig.\,\ref{FIG:8}(c), the spike in $G$ is significantly broadened and the plateau of $G$ on both
sides of the spike is reduced simultaneously with  increasing $V_b$. This is  similar to the hot-carrier
effect with increased $T$.
\medskip
\par

We now turn our attention to the experimental aspects which are related to the
preceding theoretical results. Two-terminal differential conductance measurements
were performed with an excitation voltage of $10\,\mu$V at $73\,$Hz using the Oxford
Instruments cryofree dilution refrigerator, where the device is
estimated to have an electron temperature of around $70\,$mK.
\medskip
\par

\begin{figure}[t]
\centering
\includegraphics[width=0.45\textwidth]{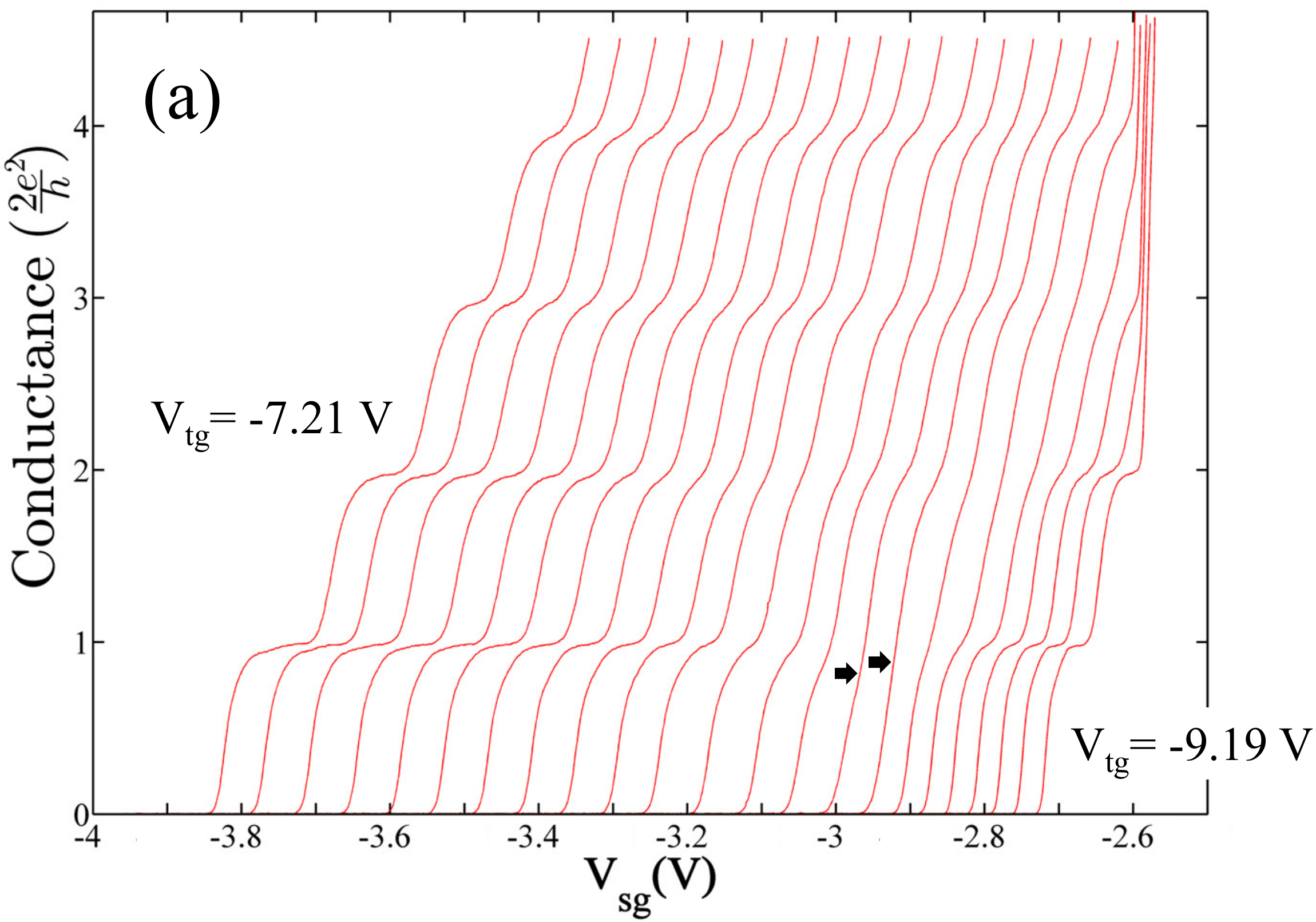}
\includegraphics[width=0.45\textwidth]{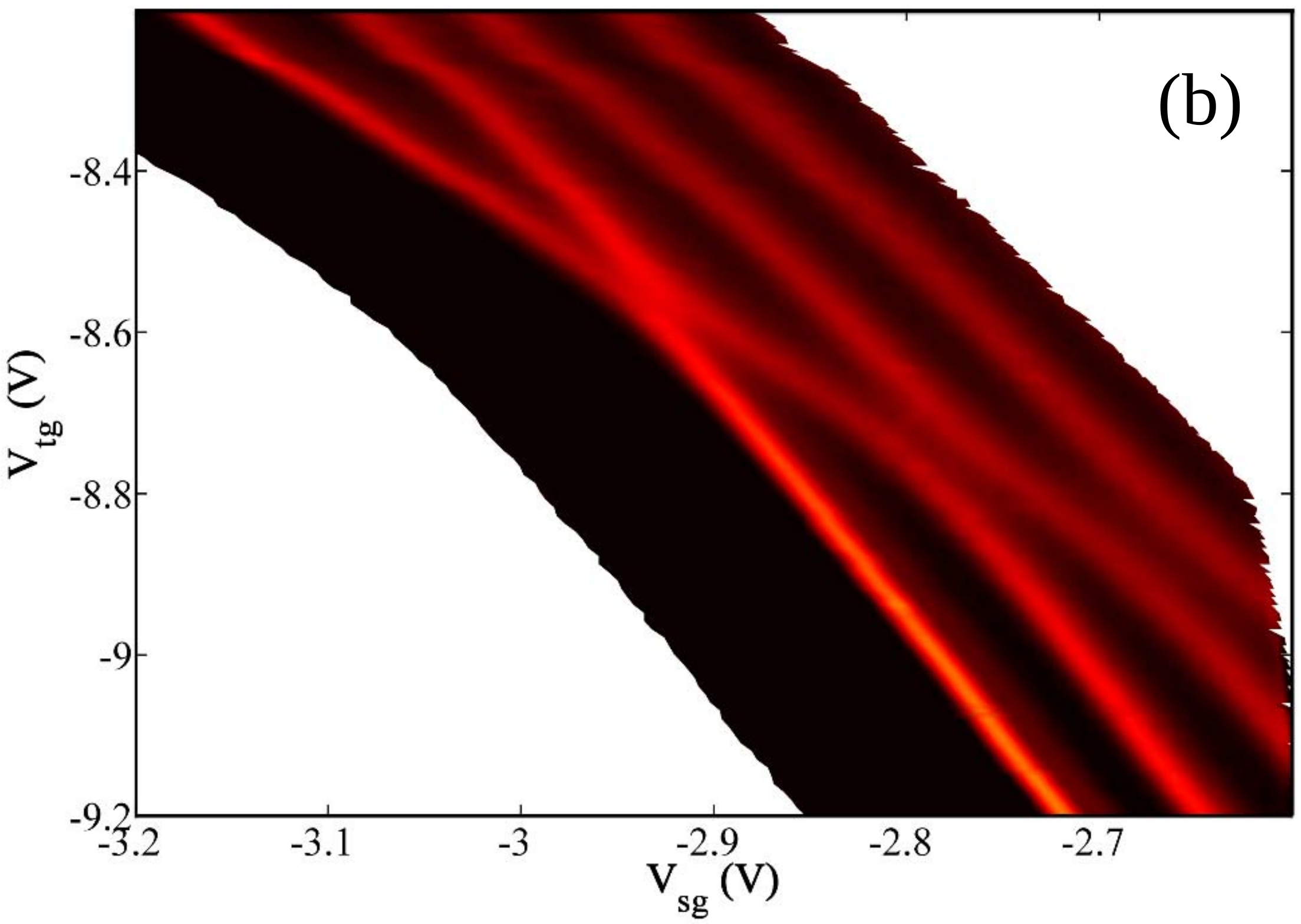}
\caption{(Color online) Plot of measured differential conductance in (a) as functions of split gate voltage $V_{\rm sg}$ for various values of top gate voltage $V_{\rm tg}$,
and in (b) the transconductance ($dG/dV_{\rm sg})$ plot of the data shown in (a). 
The confinement in (a) is controlled by making the top gate negative so that left (right) of the plot is strong (weak) confinement, where	
a direct jump to $4e^2/h$ (indicated by arrows) occurs when the confinement is weakened using a top gated, split-gate device. In addition,
the first trace in (a) on the left is taken at $V_{\rm tg}=-7.21$\,V and successive traces were plotted in steps of $90$\,mV until $V_{\rm tg}=-9.19$\,V.}
\label{FIG:9}
\end{figure}

In order to the test  the samples, a top gated, split gate device  provided additional
confinement to the  quasi-1D electrons.  This allowed us to  vary the confinement from being very
strong (zero top gate) to very weak (very negative top gate voltage). In the present study, as
shown in Fig.\,\ref{FIG:9}, the top gate voltage, $V_{\rm tg}$, is varied from $-7.21\,$V (left) to
$-9.19\,$V (right) in  steps of $90\,$mV.
\medskip
\par

Figure\ \ref{FIG:9} shows a plot of the differential conductance in (a) for the device as a function of the
split gate voltage $V_{\rm sg}$ for various values of the top gate voltage $V_{\rm tg}$, 
as well as in (b) for the transconductance ($dG/dV_{\rm sg})$ drawn from the data in (a). 
As can be seen from Fig.\,\ref{FIG:9}(a), as the confinement is reduced, the $2e^2/h$ conductance
plateau is weakened. If the confinement is further reduced, the $2e^2/h$ plateau disappears and
is replaced by a direct jump in conductance to the (rounded) $4e^2/h$ plateau at both
$V_{\rm tg}=-8.47$ and $-8.56$\,V (indicated by arrows). Eventually the first plateau at $2e^2/h$ is
 recovered on further reducing the confinement to $V_{\rm tg}=-9.19$\,V (right-most red curve).
In comparison with our calculated results presented in Figs.\,\,\ref{FIG:3} and \,\ref{FIG:4},
we find the sequence from the appearance of the $2e^2/h$ conductance plateau for small values
of ${\cal R}$.   We have obtained results for  strong confinement as well as the $4e^2/h$
conductance plateau for intermediate confinement, and again the $2e^2/h$ conductance plateau
in the weak-confinement regime which is preceded by a double-kink structure.
In addition, from Fig.\,\ref{FIG:9}(b) we know the crossing/anticrossing of the ground state and the first excited states 
depends on the confinement strength. Here, when $V_{\rm tg}$ is around $-8.6\,$V, the ground state and the first excited states cross, leading to energy reversal 
such that previous excited state becomes the new ground state, and then, the previous ground state further moves up in the energy and anticrosses with the second excited state.
This observation qualitatively agrees with the calculated results presented in Fig.\,\ref{FIG:5}.
\medskip
\par

We would like to emphasize that the appearance/disappearance/reappearance of a
conductance plateau has been qualitatively reproduced in our numerical calculations. This is
displayed in Fig.\,\ref{FIG:6}, although some non-monotonic features in our reported results are not
verified experimentally. We acknowledge that there is some non-monotonic behavior in the results of our
simulations, e.g., in Figs.\,\ref{FIG:6} and \ref{FIG:4}, preceding the onset of the first conductance
plateau which is verified by the experimental  data. However, apart from this, we do believe
that we have qualitatively reproduced a significant part of the experimentally observed
recurrence of the first conductance plateau with increasing channel width.  This is an
aim of our review, and such an observation highlights the importance of the Coulomb interaction
between electrons  after appreciably  suppressing the electron kinetic energy contribution
as the channel confinement becomes very weak.

\section{Concluding Remarks}
\label{sec:4}

The ballistic conductance for a quasi-1D channel  (quantum wire) has exhibited an interesting
behavior as functions of the electron density as well as confinement.
We demonstrated that  electron-electron interaction plays a
crucial role in our calculations in the weak confinement regime. Extensive calculations were  carried out
in regards the effects due to confinement on the conductance and its associated dependence
on the interplay between level anticrossing and crossing in quantum transport of two
interacting-electron clusters.  As shown in our numerical results, depending
on the  confinement parameter, the conductance manifests the signature of single-particle
 or interacting two-electron state behavior. This  dependence can be observed in the deviation
of the conductance from $2e^2/h$ (single-particle) to $4e^2/h$ (interacting crossing state)
and back to $2e^2/h$ (interacting anticrossing state) as a function of the width of the quantum
wire. It is interesting to observe how many-body  effects enter  the calculation of the
quantum ballistic conductance,  where the center-of-mass velocity is not affected by the
electron-electron interaction but the electron distribution is affected.
\medskip
\par

We conclude that the experimental observations qualitatively agree well with our theoretical calculations.
Furthermore, such  experimentally observed features for switching conductance plateau can be physically explained
by the interchange of the ground between   $E^{(p)}_-$ and the degenerate $E^{(p)}_3$ and $E^{(p)}_4$ and back
to $E^{(p)}_-$, which is reflected as an upward jump from $2e^2/h$ to $4e^2/h$ and followed by another step
jump from $4e^2/h$ back to $2e^2/h$ with increasing channel width.

\begin{acknowledgements}
DH would like to thank the Air Force Office of Scientific Research (AFOSR) for the support.
\end{acknowledgements}

\clearpage

\end{document}